# Chiral-Structured Superconductors $TrX_4$ (Tr = Rh, Ir; X = Ge, Si): A Platform for Mixed-Parity Pairing and Topological States

Zhenhai Yu[1†], Yunguan Ye[1†], Yuwei Zhou[2†], Chaoyang Chu[1†], Congcong Le[3†], Lin Wu[4], Jian Yuan[1], Tong Shi[5], Qingxin Dong[5], Jinggeng Zhao[6], Wei Xia[1,7], Xiangqi Liu[1], Xia Wang[1,8], Bosen Wang[5], Jinguang Cheng[5], Yanhang Ma[1*], Xianxin Wu[9*], Xiangang Wan[3,4*], Huiqiu Yuan[2*], and Yanfeng Guo[1,7*]

[1]*State Key Laboratory of Quantum Functional Materials, School of Physical Science and Technology, ShanghaiTech University, Shanghai 201210, China*

[2]*Center for Correlated Matter and School of Physics, Zhejiang University, Hangzhou 310058, China*

[3]*Hefei National Laboratory, Hefei 230088, China*

[4]*National Laboratory of Solid-State Microstructures, School of Physics, Nanjing University, Nanjing 210093, China*

[5]*Beijing National Laboratory for Condensed Matter Physics and Institute of Physics, Chinese Academy of Sciences, Beijing 100190, China*

[6]*School of Physics, Harbin Institute of Technology, Harbin 150080, China*

[7]*ShanghaiTech Laboratory for Topological Physics, ShanghaiTech University, Shanghai 201210, China*

[8]*Analytical Instrumentation Center, School of Physical Science and Technology, ShanghaiTech University, Shanghai 201210, China*

[9]*CAS Key Laboratory of Theoretical Physics, Institute of Theoretical Physics, Chinese Academy of Sciences, Beijing 100190, China*

**ABSTRACT**

**Chiral-structured superconductors, with simultaneous broken mirror and inversion symmetries, promote unconventional superconductivity through parity-mixing mechanisms. Yet a few bulk chiral-structured superconductors are known, partly due to the difficulty in directly determining their atomic-scale chirality. Here we report three chiral-structured superconductors, $RhGe_4$, $IrGe_4$, and $IrSi_4$, synthesized under high pressure, with $T_c$ values of ~ 2.6 K, 1.1 K, and 2.5 K, respectively. Using atomic-resolution Cs-corrected scanning transmission electron microscopy (STEM) combined with X-ray diffraction characterizations, we directly confirm their chiral structure (space group $P3_121$). This real-space imaging approach overcomes ambiguities in traditional diffraction-based methods. These materials exhibit type-II superconductivity, and the enhancement of spin–orbit coupling (SOC) leads to the emergence of mixed parity pairing. Calculations also reveal symmetry-protected Weyl points near the Fermi level, which is robust against the SOC. Our work not only expands the family of chiral-structured superconductors but also demonstrates the indispensable role of STEM in directly determining chiral crystal structures. These materials thus offer a clean platform to explore the interplay among structural chirality, spin-orbit coupling, mixed-parity superconductivity, and topological quantum phenomena.**

## INTRODUCTION

Chirality is a fundamental concept across physics, chemistry, and biology, known for enabling phenomena such as spin selectivity, enantioselective catalysis, and chiral drug design [1–5]. In crystalline materials, structural chirality arises from the absence of mirror planes, inversion centers, and rotational-reflection axes, resulting in a structure that cannot be superimposed on its mirror image. This symmetry breaking has sparked significant research interest, leading to discoveries of distinct physical properties in solids, including spin-selective effects [4], nonreciprocal transport [6], circularly polarized optical responses [7], and chiral topological fermions [8]. Recently, even electrical-current-induced orbital magnetization has been observed in chiral crystals [9,10]. Given such unique manifestations, understanding how structural chirality influences superconductivity represents a compelling scientific direction, yet progress has been limited by the scarcity of confirmed chiral-structured superconductors. In most known superconductors, such as conventional *s*-wave, high-$T_c$ cuprate, and iron-pnictide systems, Cooper pairing is spin-singlet, protected by inversion symmetry. In noncentrosymmetric superconductors, however, the absence of inversion symmetry lifts the momentum-degenerate at each $\boldsymbol{k}$ point, enabling antisymmetric spin-orbit coupling (ASOC) to mix spin-singlet and spin-triplet pairing components [11]. This mixing can give rise to distinctive properties, including anomalous penetration depth and upper critical fields exceeding the Pauli limit [12–15]. Despite such intriguing prospects, verified chiral superconductors, such as those with chiral *p*- or *f*-wave symmetry, remain exceptionally rare. Even widely studied candidates like $UPt_3$ [16], $Cu_xBi_2Se_3$ [17], and $Sr_2RuO_4$ [18] continue to face debate regarding their precise pairing symmetries [19].

The key experimental challenges in this field lie in single-crystal growth and reliable determination of chiral crystal structures. So far, most known chiral-structured superconductors exist only as polycrystalline samples, which limits further characterization and impedes deeper insight into their unique physical properties. Moreover, many chiral space groups exhibit only subtle differences in their diffraction patterns, making it difficult, and often impossible, to distinguish between enantiomorphic or closely related chiral phases using conventional X-ray diffraction (XRD) alone. Without direct real-space imaging or complementary local symmetry probes, chiral assignments can remain ambiguous, hindering clear structure-property correlations. When realized, chiral-structured superconductors are predicted to host a rich spectrum of exotic phenomena. If the superconducting order parameter winds in momentum space, breaking time-reversal symmetry, topological superconductivity with Majorana boundary states can emerge [20]. Additional anticipated effects include superconducting diode behavior [21], chiral topological fermions [8], and protected surface states [22]. Thus, the discovery and definitive structural characterization of chiral-structured superconductors could open new pathways toward topological quantum matter and unconventional superconducting responses.

By using the high-pressure (HP) synthesis technique, we successfully grew single crystals of chiral-structured superconductors $RhGe_4$ and $IrGe_4$, as well as polycrystalline $IrSi_4$. Combining single-crystal and powder XRD with atomic-resolution Cs-corrected scanning transmission electron microscopy (STEM), we directly and unambiguously determined their chiral crystal symmetry. Resistivity, magnetization, and specific heat measurements confirm bulk superconductivity in all three materials. As the SOC increases from $RhGe_4$ to $IrGe_4$, the superconducting transition temperature decreases, while the *p*-wave contribution to the gap symmetry grows—as evidenced by the low-temperature magnetic penetration depth measured using a tunnel diode oscillator (TDO) technique, as well as the specific heat. First-principles calculations further reveal the presence of symmetry-protected Weyl points at high-symmetry positions and lines in their electronic structures, highlighting the intertwined nature of chiral crystal symmetry, superconductivity, and topological states in these materials.

## EXPERIMENTAL METHODS

**HP Synthesis.** Polycrystalline $RhGe_4$ was previously synthesized using HP methods [23,24], but single crystals remained unavailable. In contrast, $IrGe_4$ single crystals have been successfully obtained via the Czochralski method in a tetra-arc furnace [25]. For a direct comparison between the two materials, we synthesized single crystals of both compounds under uniform conditions using the HP technique. Besides, the crystal growth of a new compound $IrSi_4$ was also tried. Starting materials Rh/Ir and Si/Ge powders were mixed in a molar ratio of (Rh/Ir): (Si/Ge) = 1: 4 in an Ar-filled glove box. The mixture was pre-compressed into a cylindrical pellet and loaded into a hexagonal boron nitride (h-BN) sample chamber, which was then placed inside a graphite heating furnace and enclosed within an MgO octahedron. Single crystals of $RhGe_4$ and $IrGe_4$ were grown under 3–5 GPa and 800–1000 °C using a Kawai-type multi-anvil apparatus. Attempts to synthesize $IrSi_4$ were carried out at 5–7 GPa and 1200–1400 °C; however, despite numerous trials, the formation of crystalline $IrSi_4$ was not achieved. Further details on the HP synthesis are provided in our previous work [26]. Rh and Ir belong to the same group VIIIB but differ in their electron configurations ($4d^8 5s^1$ for Rh versus $5d^7 6s^2$ for Ir), which leads to distinct SOC strengths. The availability of $RhGe_4$ and $IrGe_4$ enables a systematic investigation of how SOC influences superconductivity in the chiral structure.

**Compositions and Crystal Structure Characterizations.** The chemical compositions of the products were examined by energy-dispersive X-ray spectroscopy (EDX). Measurements taken at multiple points on each crystal were averaged to ensure representative stoichiometry. As exemplified by $RhGe_4$ and $IrGe_4$ (Fig. S1), the EDX results confirm a consistent atomic ratio of Rh: Ge = 1: 4 and Ir: Ge = 1: 4, respectively. Crystal structures were identified using a single-crystal X-ray diffractometer equipped with a Mo $K_\alpha$ source ($\lambda$ = 0.71073 Å) at 290 K. Structural refinements were performed with the SHELXL package implemented in Olex2 [27]. Powder X-ray diffraction (PXRD) data were collected on all samples using a Bruker D8 diffractometer with Cu $K_\alpha$ radiation ($\lambda$ = 1.5418 Å) at 300 K. Rietveld refinements of the XRD patterns were carried out with the GSAS program suite via the EXPGUI interface [28].

**Cs-corrected STEM Measurements.** Atomic-scale structural characterization was performed using Cs-corrected STEM. Single crystals were lightly crushed, and the resulting powder was dispersed in ethanol via sonication for 5 minutes. A drop of the suspension was deposited onto an ultrathin pure-carbon film and allowed to dry naturally. Selected-area electron diffraction (SAED) patterns and atomic-resolution high-angle annular dark-field (HAADF) STEM images were acquired from isolated thin flakes. All HAADF-STEM imaging was conducted at 300 kV using a cold-field-emission gun, aberration-corrected JEOL GRAND ARM 300F microscope. Atomic projections of $RhGe_4$ were visualized in the VESTA software [29].

**Physical Properties Measurements.** Electrical resistivity and specific heat were measured in a Quantum Design Physical Property Measurement System (PPMS-9) using the standard four-probe and relaxation methods, respectively. Specific-heat measurements down to 0.4 K were performed in an adiabatic relaxation mode with a $^3$He refrigerator insert, under applied magnetic fields up to 1000 Oe. DC magnetization was collected in a Quantum Design Magnetic Property Measurement System (MPMS). Both zero-field-cooled (ZFC) and field-cooled (FC) protocols were employed to fully characterize the superconducting diamagnetic response.

**TDO Measurements.** The magnetic penetration depth $\lambda(T)$ is an important parameter for characterizing the superconducting order parameter. We measured the temperature dependence of the magnetic penetration depth of $RhGe_4$ and $IrGe_4$ using a TDO-based technique in a $^3$He cryostat and a dilution refrigerator, reaching a base temperature of 0.33 K and 0.055 K, respectively. The respective operating frequencies of the TDO were approximately 7 MHz and 14 MHz in the two cryostats, with noise levels of about 0.15 Hz and 0.5 Hz [30,31]. The change of the magnetic penetration depth from the zero-temperature value, $\Delta\lambda(T) = \lambda(T) - \lambda(0)$, is proportional to the TDO resonant frequency shift $\Delta f(T) = f(T) - f(0)$, namely $\Delta\lambda(T) = G\Delta f(T)$. Here the geometric G-factor is estimated following reference [31]. The samples used in this study were cut into regular rectangular shapes with dimensions of 260×210×20 μm$^3$ for $RhGe_4$ and 340×430×100 μm$^3$ for $IrGe_4$.

**First-Principles Calculations.** First-principles calculations were performed using the projector augmented wave (PAW) method as implemented in the Vienna *Ab initio* Simulation Package (VASP) [32,33]. The Perdew–Burke–Ernzerhof (PBE) exchange-correlation functional [34] was adopted. A plane-wave kinetic-energy cutoff of 500 eV was used, and the Brillouin zone was sampled with a Monkhorst–Pack k-point mesh [35]. Based on the experimentally determined crystal structures, an 8 × 8 × 6 k-mesh was employed for all calculations.

## RESULTS AND DISSCUSSION

The crystalline phase and phase purity of $RhGe_4$, $IrGe_4$, and $IrSi_4$ were examined via PXRD. Rietveld refinements of the diffraction patterns are summarized in Figs. 1(g)–1(i) and detailed in Tables S1-S3. All three compounds crystallize in an $IrGe_4$-type chiral structure with a space group $P3_121$ (No. 152). The refined lattice parameters are $a$ = 6.1906(5) Å and $c$ = 7.8082(7) Å for $RhGe_4$, $a$ = 6.1974(13) Å and $c$ = 7.7649(18) Å for $IrGe_4$, $a$ = 5.9629(3) Å and $c$ = 7.4777(3) Å for $IrSi_4$, respectively. These results are consistent with previous reports [24,25,34]. For $IrSi_4$, a minor non-superconducting metal impurity phase of $IrSi_3$ (~ 10 % by weight) was detected.

In this chiral structure, the Rh/Ir atoms occupy the 3a Wyckoff position, while the Ge/Si atoms are distributed over three inequivalent Wyckoff positions: Ge(1)/Si(1) at 3b, Ge(2)/Si(2) at 3b, and Ge(3)/Si(3) at 6c [Fig. 1(d)]. The refined structural parameters are listed in Table 1. The chiral crystal structures are illustrated in detail in Fig. 1(e). Each Rh/Ir atom is coordinated by eight Ge/Si atoms, forming a distorted dodecahedral [$RhGe_8$], [$IrGe_8$] or [$IrSi_8$] polyhedron [Fig. 1(a)]. Within a unit cell, only one such polyhedron shares an edge with its neighbor (highlighted by an ellipse in Fig. 1(b)); the remaining five are connected via vertex-sharing. These polyhedra arrange into a layered architecture along the stacking direction [Fig. 1(c)], which has been suggested to favor two-dimensional conduction and may promote superconductivity [11]. The schematic crystal structure highlighting the three inequivalent Wyckoff positions is shown in Fig. 1(d). Fig. 1(e) depicts the stacking of Rh/Ir and Si/Ge atoms along the (00-1) and (001) directions, where the chirality of the structure is manifested by the left- or right-handed configuration of the ab-plane. Finally, the arrangement of dumbbell-like Ge–Ge and Si–Si bonds in $RhGe_4$, $IrGe_4$ and $IrSi_4$ is displayed in Fig. 1(f).

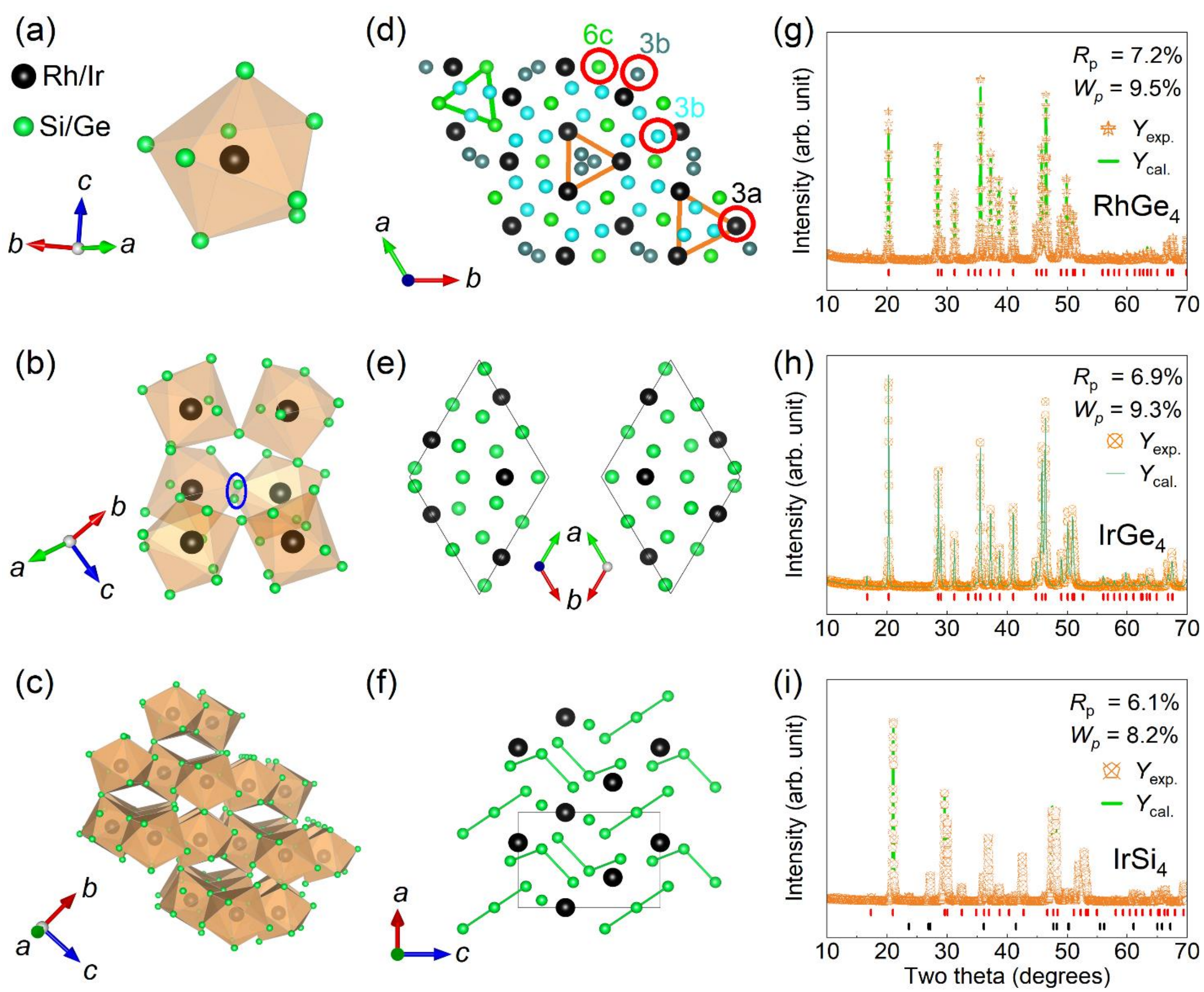

**FIG. 1.** (a) Distorted [$RhGe_8$], [$IrGe_8$] and [$IrSi_8$] dodecahedral coordination polyhedra. (b) Connectivity of polyhedra via vertex-sharing (dominant) and one edge-sharing connection (highlighted by an ellipse). (c) Layered arrangement of [$RhGe_8$]/[$IrGe_8$]/[$IrSi_8$] polyhedra along the *b*-axis. (d) Schematic crystal structures of $RhGe_4$, $IrGe_4$ and $IrSi_4$ in space group $P3_121$, showing the three inequivalent Wyckoff positions. (e) Projection of the chiral crystal structures viewed along the (00-1) and (001) directions, illustrating the left-/right-handed configuration in the *ab*-plane. (f) Arrangement of dumbbell-like Ge–Ge and Si–Si covalent bonds. (g)–(i) Rietveld refinement profiles for PXRD data of $RhGe_4$, $IrGe_4$ and $IrSi_4$,

respectively. Vertical tick marks indicate the allowed Bragg reflections for $RhGe_4$ and $IrGe_4$. For $IrSi_4$, the majority phase is indicated by the main set of ticks, while a minor impurity phase of $IrSi_3$ is also marked (denoted by asterisks).

**Table I.** Selected structural parameters of Rietveld refinements for $RhGe_4$, $IrGe_4$ and $IrSi_4$.

| $RhGe_4$, SG: $P3_121$, $a$ = 6.1906(5) Å and $c$ = 7.8082(7) Å | | | | |
|---|---|---|---|---|
| Atom | WP | $x$ | $y$ | $z$ |
| Rh | 3a | 0.3175 | 0 | 0.3333 |
| Ge(1) | 3b | 0.0769 | 0 | 0.8333 |
| Ge(2) | 3b | 0.6159 | 0 | 0.8333 |
| Ge(3) | 6c | 0.5116 | -0.2655 | 0.2785 |
| $IrGe_4$, SG: $P3_121$, $a$ = 6.1974(13) Å and $c$ = 7.7649(18) Å | | | | |
| Atom | WP | $x$ | $y$ | $z$ |
| Ir | 3a | 0.3177 | 0 | 0.3333 |
| Ge(1) | 3b | 0.0773 | 0 | 0.8333 |
| Ge(2) | 3b | 0.6169 | 0 | 0.8333 |
| Ge(3) | 6c | 0.2119 | 0.7339 | 0.3855 |
| $IrSi_4$, SG: $P3_121$, $a$ = 5.9629(3) Å and $c$ = 7.4777(3) Å | | | | |
| Atom | WP | $x$ | $y$ | $z$ |
| Ir | 3a | 0.3147 | 0 | 0.3333 |
| Si(1) | 3b | 0.0701 | 0 | 0.8333 |
| Si(2) | 3b | 0.6073 | 0 | 0.8333 |
| Si(3) | 6c | 0.4866 | -0.2704 | 0.2566 |

Selected Rh–Ge and Ir–Si bond distances are listed in Table S4, revealing four distinct bond lengths that correspond to the distorted [$RhGe_8$], [$IrGe_8$] and [$IrSi_8$] polyhedra. Both $RhGe_4$, $IrGe_4$ and $IrSi_4$ also feature three sets of Ge–Ge and Si–Si distances, as summarized in Table S4. The shortest contacts, 2.57(1) Å for Ge1–Ge3 and 2.43(5) Å for Si1–Si3, are slightly longer than the respective covalent bond lengths in diamond-structure Ge and Si (2.45 and 2.36 Å), and are identified as covalent Ge–Ge and Si–Si bonds. The resulting dumbbell-like Ge–Ge and Si–Si arrangements are illustrated in Fig. 1(f). Similar dumbbell motifs have been reported in other superconductors such as $SrGe_3$ [37] and $MSi_3$ (M = Ca, Y, Lu) [38].

However, conventional SXRD and PXRD alone cannot unambiguously assign the chiral space groups of $RhGe_4$, $IrGe_4$, and $IrSi_4$ nanocrystals, because space groups $P3_121$ and $P3_221$ (No. 154) share identical Wyckoff position configurations (3a, 3b, and 6c). From a crystallographic perspective, assigning $P3_121$ or $P3_221$ to these compounds based on SXRD and PXRD data could also be well justified. To resolve this ambiguity, we employed Cs-aberration corrected STEM, a technique proven effective for chiral structure determination [39]. High-resolution STEM imaging was performed on electron-transparent lamellae prepared from bulk $RhGe_4$ single crystals. By acquiring atomic-resolution HAADF images along the [010], [-110], and [120] zone axes, we compared the experimental projections directly with structural models for both candidate space groups. Since different space groups necessarily produce distinct atomic arrangements when viewed along specific crystallographic directions, this comparative approach allows clear discrimination between $P3_121$ and $P3_221$, as illustrated by the atomic projections along the [010], [-110], and [120] zone axes [Figs. 2(a)–2(c)]. In the HAADF-STEM images, the brightest dots correspond to Rh columns (contrast approximately scales with $Z^2$ where $Z$ denotes the atomic number), while Ge columns are marked by green spheres in the overlaid structural models. The experimental images along all three zone axes [Figs. 2(d)–2(k)] consistently match the simulated projections for the $P3_121$ structure. This real-space imaging definitively assigns the chiral space group of $RhGe_4$ as $P3_121$, demonstrating the indispensable role of atomic-resolution STEM imaging in determining the absolute structure of non-centrosymmetric crystals. The chiral structure ($P3_121$) of $IrGe_4$ was also revealed by STEM imaging (Figs. S2a and 2b).

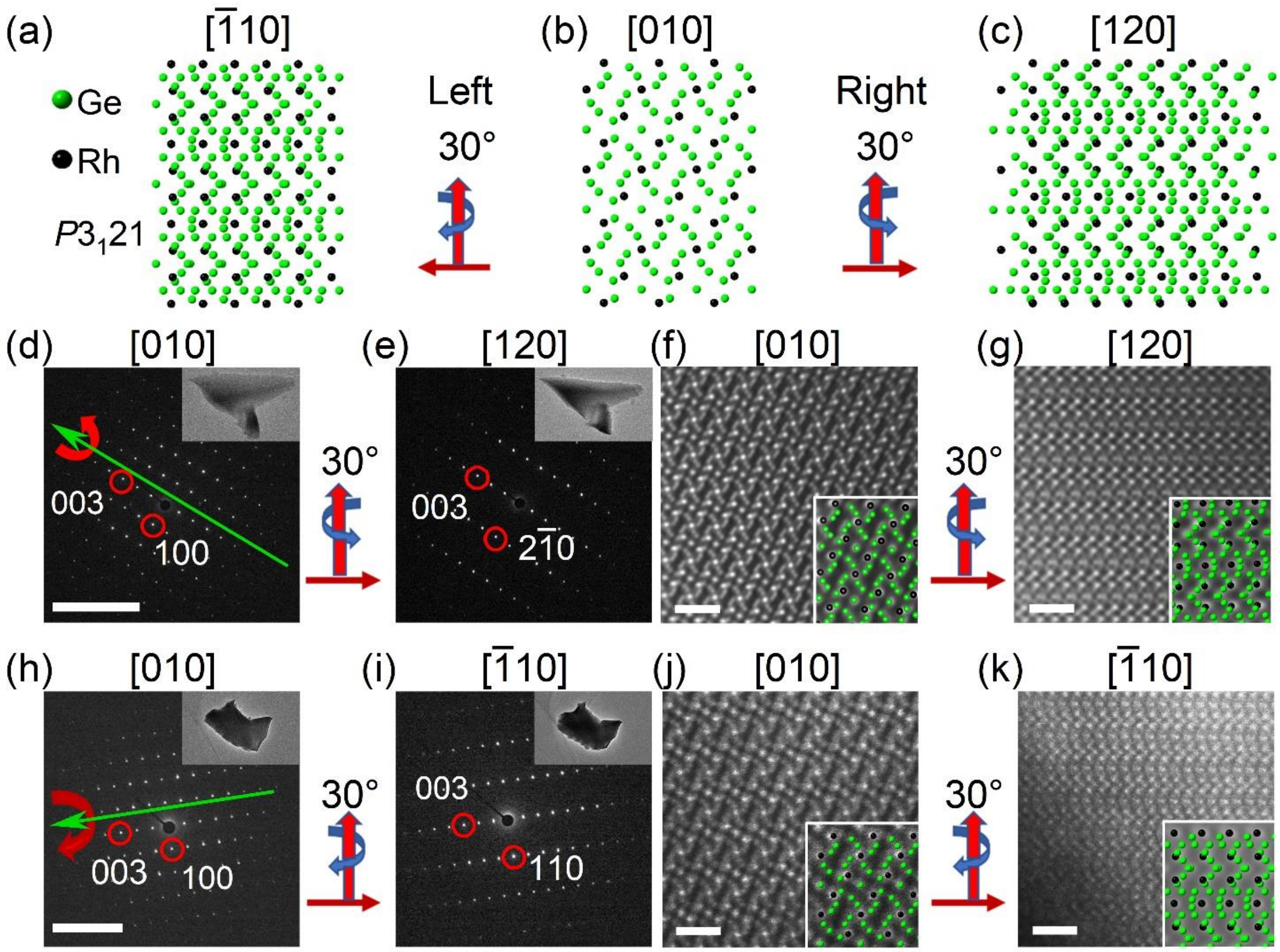

**FIG. 2.** (a)-(c) Projections of the $P3_121$ structural models of $RhGe_4$ with right-handedness along different orientations with a tilt angle of 30º. (d)-(k) STEM-ADF images of the $RhGe_4$ single crystal along [010], [-110] and [120] zone axes with a right/left tilt angle of 30º (insets are simulated and $p1$ symmetry-averaged images overlaid with structural models, with black and green spheres representing Rh and Ge atoms, respectively.), respectively. The scale bars for (d)-(e), (h)-(i), and (f)-(g), (j)-(k) are 5 nm$^{-1}$ and 1 nm, respectively.

Bulk superconductivity in $RhGe_4$, $IrGe_4$, and $IrSi_4$ was first examined through electrical transport measurements (Fig. S3). The resistivity $\rho(T)$ of all three compounds decreases gradually upon cooling, with clear superconducting transitions observed below $T_c \approx 2.6$ K ($RhGe_4$), 1.1 K ($IrGe_4$), and 2.5 K ($IrSi_4$), as shown in detail near $T_c$ in Fig. 3(a). Because the $IrSi_4$ sample contained about 10 % $IrSi_3$ impurity (Fig. 1(i)), it was essential to verify the origin of its superconductivity. Previous transport studies on pure $IrSi_3$ show normal metallic behavior down to 1.8 K with no signs of superconductivity [40]. Therefore, the observed superconducting transition in our sample is an intrinsic property of $IrSi_4$, not due to the impurity phase.

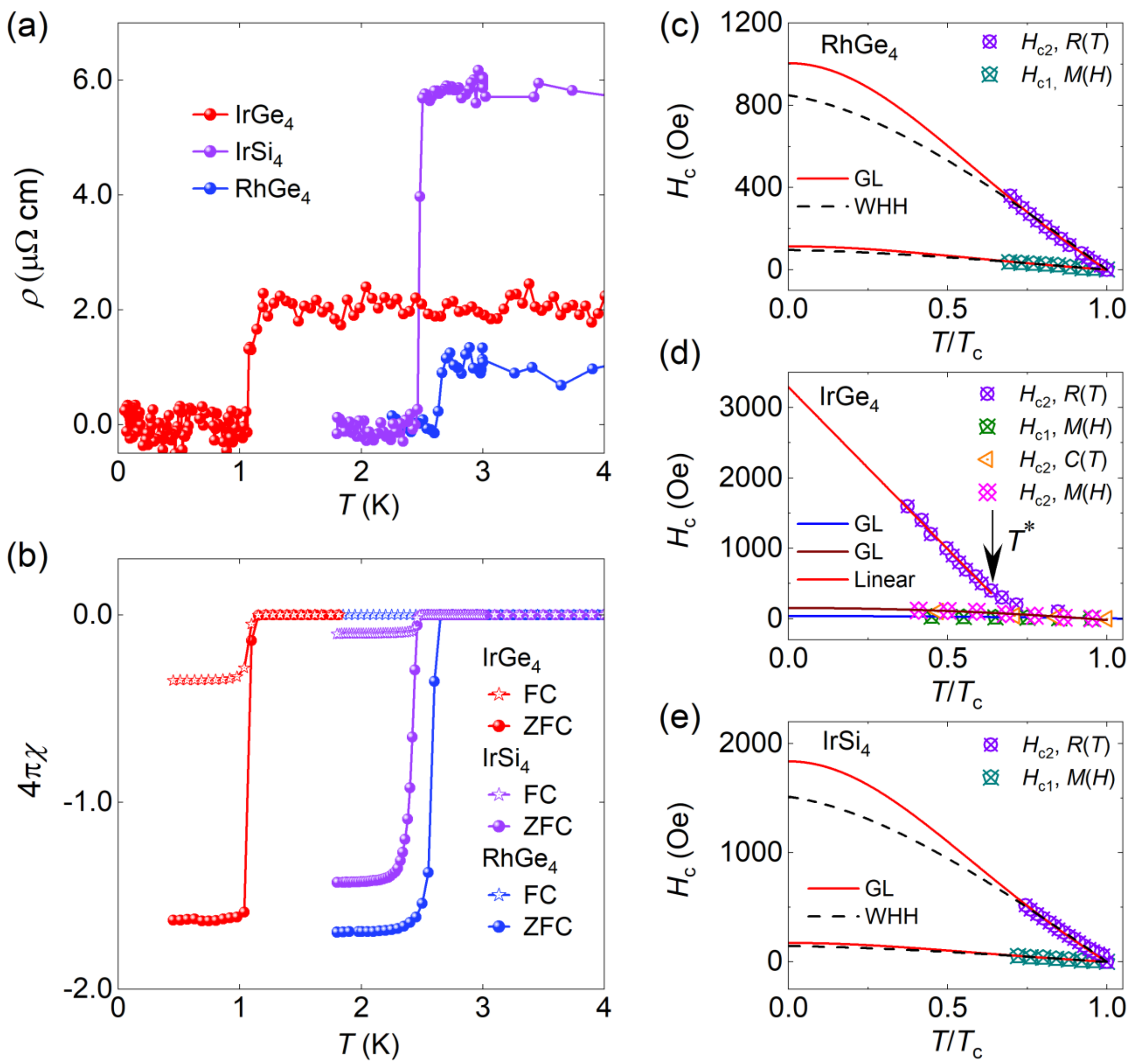

**FIG. 3.** (a) Temperature-dependent resistivity $\rho(T)$ of $RhGe_4$, $IrGe_4$, and $IrSi_4$ near the superconducting transition. (b) Magnetic susceptibility of $RhGe_4$, $IrGe_4$, and $IrSi_4$ as a function of temperature. (c)–(e) Upper critical field $H_{c2}$ and lower critical field $H_{c1}$ versus the reduced temperature $T/T_c$. $H_{c2}$ is defined from the onset transition temperature $T_c^{onset}$. For $IrGe_4$, $H_{c1}$ is extracted from $M(H)$ measurements, while $H_{c2}$ is determined from combined $\rho(T)$, $C_p(T)$, and $M(H)$ data; $T^*$ denotes the transition point in the $H_{c2}(T/T_c)$ dependence. For $RhGe_4$ and $IrSi_4$, $H_{c1}$ and $H_{c2}$ are derived from $M(H)$ and $\rho(T)$, respectively. Fits based on the Ginzburg–Landau (GL, red solid line) and Werthamer–Helfand–Hohenberg (WHH, black dashed line) models are also shown.

Figs. S4(a)–S4(c) show the resistivity of $RhGe_4$, $IrGe_4$, and $IrSi_4$ under different magnetic fields. The $T_c$ decreases with increasing field. By identifying the onset points ($T_c^{onset}$) at which $\rho$ drops from the normal state, we construct the upper-critical-field $H_{c2}$–$T$ phase diagrams, presented in Figs. 3(c)–3(e). The data are fitted with the Ginzburg-Landau (GL) equation:

$$H_{c2}(T) = H_{c2}(0)\frac{1-t^2}{1+t^2},$$

where $t = T/T_c$. The GL fits are shown as red solid lines in Figs. 3(c)–3(e). For $RhGe_4$ and $IrSi_4$, the Werthamer-Helfand-Hohenberg (WHH) model [41] is also applied (black dashed lines). Based on the $T_c^{onset}$ criterion, the zero-temperature $H_{c2}(0)$ values from both models are ~ 850 Oe (WHH) and ~ 1000 Oe (GL) for $RhGe_4$, and ~ 1500 Oe (WHH) and ~1850 Oe (GL) for $IrSi_4$. For $IrGe_4$, the $H_{c2}$ extracted from resistivity is notably higher than those obtained from specific heat and magnetization measurements. The low-temperature portion of its $H_{c2}$–$T$ curve could not be described by either the GL or WHH model. Therefore, a linear fit was used in that region, with the crossover between nonlinear and linear behavior marked as $T^*$ in Fig. 3(d). Below $T^*$, the curve exhibits an anomalous low-temperature divergence without saturation—a feature also reported in other chiral (e.g., $NbGe_2$ [42]) and non-centrosymmetric (e.g., $LaRhSi_3$ [43]) superconductors, as well as in the topological-superconductor candidate $PdTe_2$ [44]. We suggest that our data exclude mechanisms for the effect associated with surface superconductivity, implying that topologically protected superconducting surface states may be involved [45,46].

To exclude filamentary superconductivity, the Meissner effect was probed via low-field (5–10 Oe) magnetic susceptibility, shown in Fig. 3(b). Strong diamagnetic signals in the zero-field-cooled (ZFC) curves confirm bulk superconductivity in all three compounds. The $T_c$ values, determined from the intersection of the steepest superconducting slope with the extrapolated normal-state susceptibility, are ~ 2.6 K ($RhGe_4$), ~ 1.1 K ($IrGe_4$), and ~ 2.5 K ($IrSi_4$). After demagnetization correction [47], the ZFC shielding fraction reaches ~ 90 %. Magnetization versus field curves at various sub-$T_c$ temperatures [Figs. S4(d)–S4(f)] further indicate conventional type-II behavior. The lower critical field $H_{c1}$, extracted from these $M(H)$ isotherms, is plotted in Figs. 3(c)–3(e). The $T_c$ values obtained from diamagnetic response are consistent with those from resistivity.

Specific heat measurements [Figs. 4(a)–4(c)] provide additional evidence for bulk superconductivity. A clear jump near $T_c$ is suppressed and shifts to lower temperatures with increasing field [Figs. S5(a)–S5(c)]; the transition is fully suppressed at 500 Oe ($RhGe_4$), 100 Oe ($IrGe_4$), and 1000 Oe ($IrSi_4$). In the normal state ($H = 0$), the data were fitted to $C_p/T = \gamma + \beta T^2$ [Figs. S5(d)–S5(f)], yielding: $\gamma$ = 5.99(5) mJ mol$^{-1}$ K$^{-2}$, $\beta$ = 0.389(3) mJ mol$^{-1}$ K$^{-4}$, $\Theta_D$ = 292.1 K for $RhGe_4$, $\gamma$ = 4.26(2) mJ mol$^{-1}$ K$^{-2}$, $\beta$ = 0.212(4) mJ mol$^{-1}$ K$^{-4}$, $\Theta_D$ = 357.8 K for $IrGe_4$, and $\gamma$ = 4.39(3) mJ mol$^{-1}$ K$^{-2}$, $\beta$ = 0.268(1) mJ mol$^{-1}$ K$^{-4}$, $\Theta_D$ = 330.7 K for $IrSi_4$. $\Theta_D = [12\pi^4 nR/(5\beta)]^{1/3}$ ($n = 5$ atoms per formula unit, $R$ the gas constant). Using the inverted McMillan formula [48] with the Coulomb pseudopotential $\mu^* = 0.05 – 0.2$, the electron–phonon coupling constants ($\lambda_{ep}$) are estimated as $\lambda_{ep} = 0.40 – 0.69$ for $RhGe_4$, $\lambda_{ep} = 0.32 – 0.58$ for $IrGe_4$, and $\lambda_{ep} = 0.38 – 0.67$ for $IrSi_4$. These values indicate weak-to-moderate coupling superconductivity. The electronic density of states at the Fermi level, $N(E_F) = 3\gamma/[\pi^2 k_B^2(1+\lambda_{ep})]$, where $k_B$ is the Boltzmann constant, is calculated as 1.82 states eV$^{-1}$ for $RhGe_4$, 1.26 states eV$^{-1}$ for $IrGe_4$, and 1.35 states eV$^{-1}$ for $IrSi_4$.

The lattice contribution $\beta T^3$ was subtracted from the total specific heat $C_p(T)$ to isolate the electronic part $C_{el}$. The temperature dependence of $C_{el}/T$ is displayed in Figs. S5(g)-S5(i). The derived normalized specific-heat jumps $\Delta C_{el}/\gamma T_c$ are 1.42 for $RhGe_4$, 1.35 for $IrGe_4$, and 1.46 for $IrSi_4$, all close to the BCS weak-coupling value of 1.43. In the superconducting state, the low-temperature $C_{el}(T)$ data were fitted with the phenomenological expression

$$C_{el}(T) = \gamma_{res}T + A\exp\left(-\frac{\Delta}{k_B T}\right),$$

where $\gamma_{res}T$ accounts for a residual electronic contribution from impurities, and the second term describes the exponential decay characteristic of an *s*-wave gap.

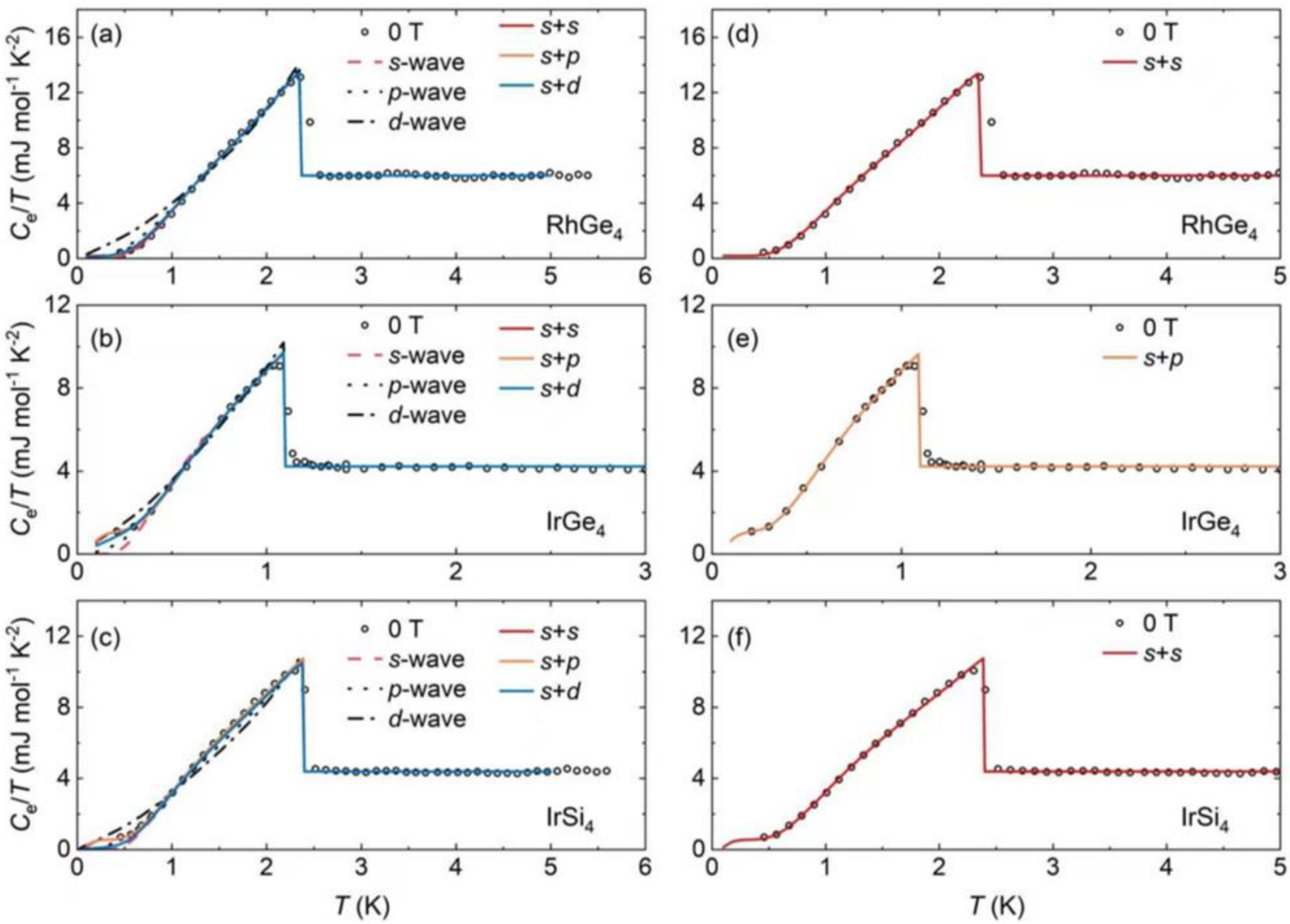

**FIG. 4.** Further characterization of the superconductivity. (a)-(c) $C_{el}/T$ versus $T$ as well as the fitting results with several pairing waves. (d)-(f) The best fitting results for panels were analyzed using '$s + s$', '$s + p$', and '$s + s$' pairing, respectively.

To probe the influence of chiral structure on superconductivity, the $C_{el}/T$ data of $RhGe_4$, $IrGe_4$, and $IrSi_4$ were analyzed using several gap-function models [14], as shown in Figs. 4(a)–4(c). The excellent agreement between experiment and theory (Figs. 4(d)–4(f)) supports a mixed '$s + s$'-wave pairing model for $RhGe_4$, whereas an '$s + p$'-wave model is favored for $IrGe_4$, with the *p*-wave component estimated to be approximately 15.7%. For $IrSi_4$, although both '$s + s$'-wave and '$s + p$'-wave models can describe the data reasonably well, the extracted gap magnitude is rather small, rendering the physical interpretation less meaningful. The fitting results of specific heat are summarized in Table S5. Both $RhGe_4$ and $IrSi_4$ exhibit dominant singlet components, with rather small triplet admixtures. This progressive enhancement likely indicates that the superconducting order parameter in these compounds contains both spin-singlet and spin-triplet components. The pronounced $p$-wave contribution in $IrGe_4$ arises directly from the strong SOC inherent to these chiral crystals. SOC not only lifts spin degeneracy but also mixes parity, so that parity is no longer a good quantum number and mixed-parity pairing is permitted. Whereas most intermetallic superconductors exhibit conventional $s$-wave pairing, the observed enhancement of the triplet component scales with the increasing SOC strength from $RhGe_4$ to $IrGe_4$, indicating that SOC is the key mechanism mediating the interplay between structural chirality and unconventional superconducting pairing symmetry. It should be noted, however, that direct measurements of the $p$-wave component using more sensitive techniques would provide more definitive conclusions than those derived from specific heat data alone.

To further investigate the superconducting order parameters of $RhGe_4$ and $IrGe_4$, we measured the magnetic penetration depth shifts $\Delta\lambda(T)$ using the TDO technique. The temperature-dependent frequency shifts from the zero-temperature value $\Delta f(T)$ for $RhGe_4$ and $IrGe_4$ are shown in the insets of Figs. 5(a) and 5(b), respectively. Both samples exhibit a sharp change at the superconducting transition, with $T_c$ values of 2.6 K ($RhGe_4$) and 1.1 K ($IrGe_4$). These $T_c$ values are consistent with those obtained from $\rho(T)$, $M(T)$, and $C(T)$ measurements. The corresponding penetration depth shifts $\Delta\lambda(T)$ were derived from $\Delta f(T)$ using $\Delta\lambda(T) = G\Delta f(T)$ and are displayed in Figs. 5(a) and 5(b). Both samples S1 and S2 synthesized in the same batch show consistent behavior, confirming the reproducibility and reliability of the measurements.

For an isotropic $s$-wave superconductor, the change of penetration depth $\Delta\lambda(T)$ at temperatures $T << T_c$ can be described by

$$\Delta\lambda(T) = \lambda(0)\sqrt{\frac{\pi\Delta(0)}{2k_BT}}\exp\left(-\frac{\Delta(0)}{k_BT}\right),$$

where $\Delta(0)$ is the superconducting gap magnitude at zero temperature. This equation is used to fit the data below $T_c/3$ for both samples. The fitted superconducting gaps $\Delta(0)$ for $RhGe_4$ and $IrGe_4$ are 1.11 $k_BT_c$ and 0.71 $k_BT_c$, respectively—both significantly smaller than the weak-coupling BCS value of 1.76 $k_BT_c$. These fitting results suggest that the superconducting gaps in the two samples may exhibit multiband features or anisotropy. To further characterize the gap structure, a power-law function ($\sim T^n$) was fitted to the data below $T_c/3$, yielding exponents $n$ of 3.5(2) for $RhGe_4$ and 2.3(1) for $IrGe_4$ (dark blue dashed lines in Figs. 5(a) and 5(b)). For $RhGe_4$, the $n = 3.5$ clearly rules out the presence of line nodes ($n = 1$) and point nodes ($n = 2$) in clean-limit superconductors, indicating a nodeless superconducting gap. In contrast, the exponent of 2.3 for $IrGe_4$ is close to $T^2$ behavior, leaving the gap structure less conclusive from power-law fitting alone.

To better resolve the superconducting gap structure, we derived the superfluid density $\rho_s(T)$ from $\Delta\lambda(T)$ using $\rho_s(T) = [\lambda(0)/(\lambda(0) + \Delta\lambda(T))]$ [31]. The zero-temperature penetration depths $\lambda(0)$ were calculated from the upper and lower critical fields, yielding values of 105 nm for $RhGe_4$ and 237 nm for $IrGe_4$. The resulting $\rho_s(T)$ are shown in Figs. 5(c) and 5(d). In the clean limit, the superfluid density can be expressed as:

$$\rho_s(T) = 1 + 2 < \int_{\Delta_k}^{+\infty} \frac{EdE}{\sqrt{E^2-\Delta_k^2}}\frac{\partial f}{\partial E} >_{FS} ,$$

where $f(E,T) = 1/(1 + e^{E/k_BT})$ is the Fermi-Dirac distribution function, $\langle \dots \rangle_{FS}$ denotes an average over the Fermi surface, and $\Delta_{\boldsymbol{k}}$ is the superconducting gap structure, which can be separated into momentum-dependent and temperature-dependent parts as:

$$\Delta_k = g_k(\theta,\phi)\Delta(T).$$

Here, $\Delta(T)$ describes the temperature dependence of the gap, often expressed using an empirical formula [49]:

$$\Delta(T) = \Delta(0)\tanh\left\{1.82\left[1.018\left(\frac{T_c}{T}-1\right)\right]^{0.51}\right\},$$

where $g_k(\theta,\phi)$ represents the angular dependence of the gap structure. For $RhGe_4$ (Fig. 5(c)), the superfluid density is well described by either a single-gap *s*-wave model ($g_k$ = 1) or a two-gap *s*-wave model with a minor contribution from the smaller gap (see SI for detailed information). In contrast, the *p*-wave model ($g_k$ = sin$\theta$) shows significant deviation at low temperatures. The fitting results are consistent with the specific heat analysis.

For $IrGe_4$ (Fig. 5(d)), the single-gap *s*-wave model deviates significantly from the experimental data, even with the optimal fitting parameter $\Delta(0) = 2.3\ k_BT_c$ (orange short dash line). The inconsistency with the penetration depth analysis persists when a two-gap *s*-wave model is used. In contrast, the *p*-wave model provides a much better fit to the data. Accordingly, $\rho_s(T)$ of $IrGe_4$ was further analyzed using *s*+*p* model and a model based on two spin-split bands [50]. Both models explain the data well across the entire temperature range (olive dashed-dotted and red solid lines in Figs. 5(d)). The underlying physical picture of the latter is as follows: due to the lack of inversion symmetry, the system exhibits ASOC, which splits the Fermi surface into two spin-split bands (labeled I and II). On these bands, the superconducting order parameter is a mixture of spin-singlet ($\psi$) and spin-triplet ($d$) components: $\Delta_\pm(\boldsymbol{k}) = \psi \pm d|\mathbf{g}_{\boldsymbol{k}}|$, where $\mathbf{g}_{\boldsymbol{k}}$ is the ASOC coupling vector determined by crystal symmetry. The model introduces two key microscopic parameters: $\delta$, the relative density-of-states difference between the two bands, and $v = \psi/d$, the singlet-to-triplet ratio. The optimal fit parameters, $\delta = 0.61$ and $\nu = 0.158$, reveal the dominant spin-triplet character and a significant difference in the density of states (DOS) between the two bands in $IrGe_4$. Specifically, the weight factors for the two bands are given by $C_I = 1 + \delta$ and $C_{II} = 1 - \delta$, giving $C_I = 1.61$ and $C_{II} = 0.39$. Thus, the nodal band II contributes only about 20% of the total DOS. Consequently, the power-law behavior arising from the nodal band is masked by the exponential behavior from the fully gapped band I, leading to a weak power-law response of the total superfluid density at low temperatures. The detailed superfluid density fitting results of $IrGe_4$ and $RhGe_4$ were shown in Table S6.

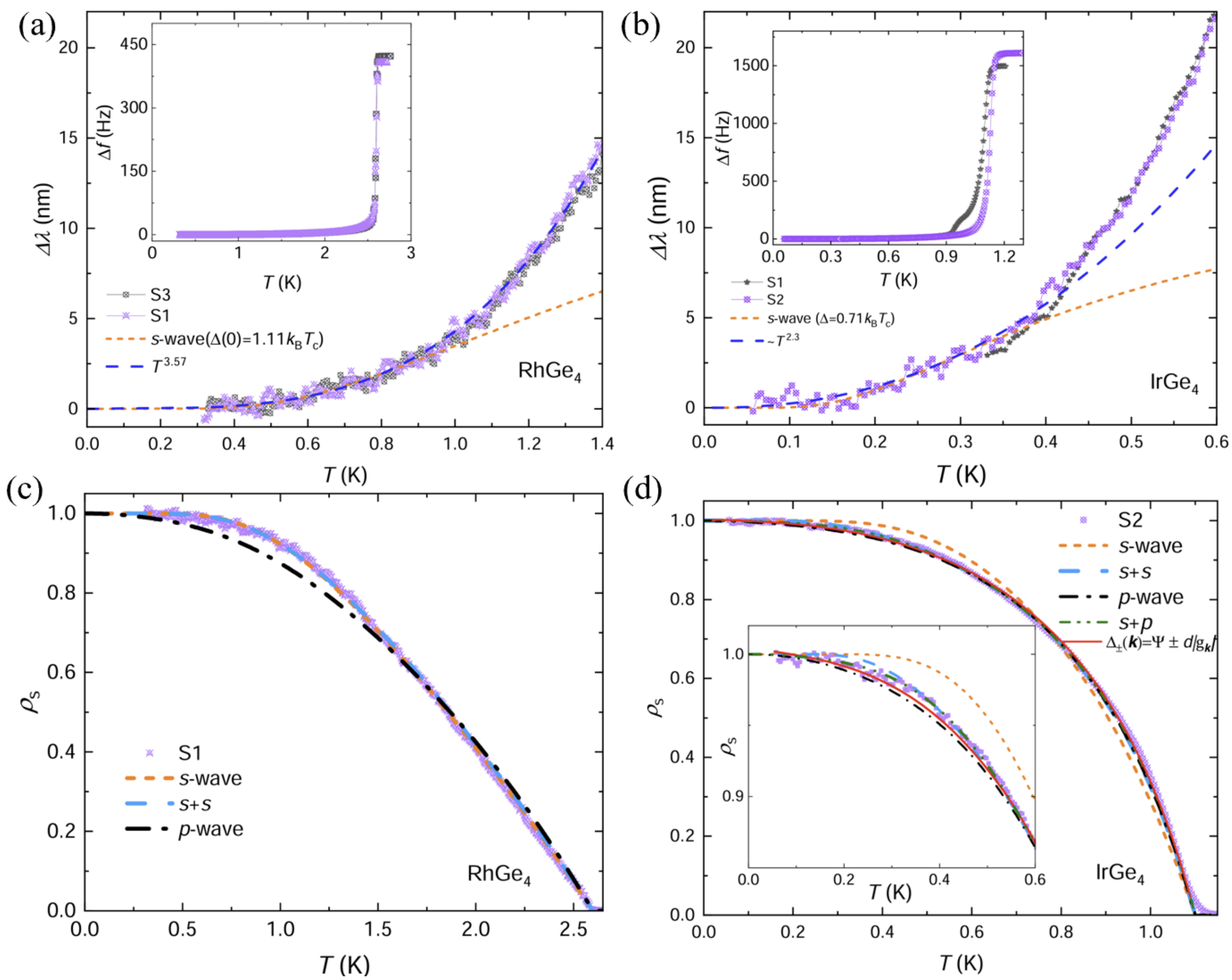

**FIG. 5.** Temperature dependence of the low-temperature $\Delta\lambda(T)$ for (a) $RhGe_4$ and (b) $IrGe_4$. The orange short dash and dark blue dashed lines represent fits using the single-band *s*-wave model and the power-law function $\sim T^n$, respectively. Insets in (a) and (b) show the full-temperature-range $\Delta f(T)$. The normalized $\rho_s(T)$ for (c) $RhGe_4$ and (d) $IrGe_4$. The orange short dash, blue dashed, black dash-dotted, olive dash-dot-dot and red solid lines correspond to fits using the *s*-wave model, two-band $\alpha$ model, *p*-wave model, *s*+*p* model and the model based on two spin-split bands, respectively.

Based on the results from both techniques, a notable difference in the order parameter between $RhGe_4$ and $IrGe_4$ is revealed. For $RhGe_4$, both specific heat and superfluid density are well described by a fully gapped spin-singlet state. The excellent agreement with the single-gap *s*-wave or an *s*+*s* two-band model, combined with a clear deviation from the *p*-wave model, indicates that the triplet component in the mixed pairing state is negligibly small in $RhGe_4$. In contrast, for $IrGe_4$, neither specific heat nor superfluid density can be described by a simple isotropic single-gap *s*-wave model. Instead, an *s*+*p* model provides a significantly better fit to the experimental data. It is worth noting that the fitted parameters obtained from the *s*+*p* model exhibit slight discrepancies between the specific heat and superfluid density measurements. We attribute this to the following two reasons. First, the lowest accessible temperatures differ between the two techniques: the superfluid density measurement reaches as low as 55 mK using a dilution refrigerator, whereas the specific heat measurement only goes down to 0.4 K using a $^3He$ refrigerator, and therefore the former are more sensitive to low energy excitations. Second, the weight parameters in the specific heat fitting reflect the DOS, while those in the superfluid density fitting also involve the Fermi velocity.

Furthermore, the superfluid density of $IrGe_4$ can be well described by a model based on two spin-split bands, consistent with de Haas–van Alphen measurements showing that the main Fermi surface of $IrGe_4$ is split into two spin-polarized bands due to ASOC [25]. These results indicate that both experimental approaches consistently point to the same conclusion: the larger spin-triplet component in $IrGe_4$ compared to $RhGe_4$ can be attributed to the increase in atomic number upon substituting Rh with Ir, which enhances the strength of ASOC.

To date, the interplay between structural chirality and superconducting properties remains poorly understood, due to the limited number of experimentally confirmed chiral superconductors and the scarcity of corresponding theoretical studies. In Table S7, we compile the crystal structure and key superconducting parameters reported in the literature for known chiral-structured superconductors, along with our synthesized compounds $RhGe_4$, $IrGe_4$, and $IrSi_4$. First-principles calculations reveal that $RhGe_4$, $IrGe_4$, and $IrSi_4$ share similar electronic band structures. We therefore select $IrSi_4$ as a representative system for detailed analysis. Fig. 6(a) displays the band structure of $IrSi_4$ along high-symmetry paths, confirming its metallic character. Notably, three distinct types of symmetry-protected Weyl points are identified at high-symmetry points and along high-symmetry lines, as illustrated in Fig. 6(b). For example, the Weyl points along the Γ–A line are protected by the $C_{3z}$ screw-rotational symmetry [51]. Similarly, the band crossings between the $J_z = +½$ and $J_z = –½$ states along the A–H line belong to different irreducible representations of the $C_2$ rotational symmetry, ensuring their topological robustness. By applying time-reversal symmetry and the $C_{3z}$ rotational symmetry, these three types of Weyl points can be extended throughout the entire Brillouin zone, generating a network of chiral fermions as mapped in Fig. 6(c). The 3D Fermi surface of $IrSi_4$ is calculated and presented in Fig. 6(d). The bands give rise to quasi-1D sheets oriented perpendicular to the $k_z$ direction, which is consistent with the observation of strong resistivity anisotropy. In addition to these quasi-1D sheets, small, closed electron-like pockets centered at the L points are also identified.

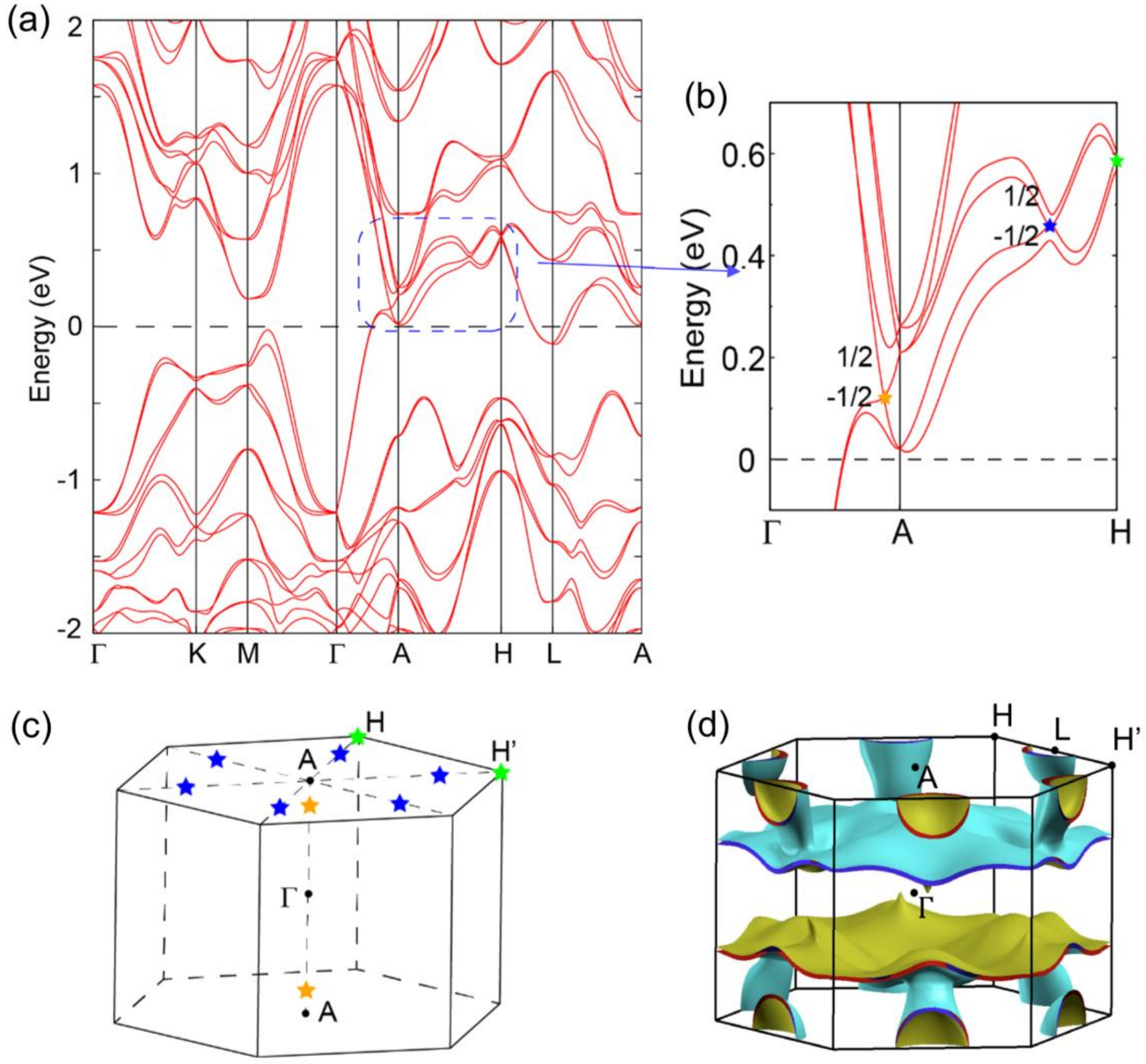

**FIG. 6.** First-principles calculations for $IrSi_4$. (a) Band structure with spin-orbital coupling. (b) Weyl points along high symmetry points and lines, highlighting their positions by stars. (c) Distribution of Weyl points in the BZ. The orange, blue and green stars denote three distinct types of Weyl points, located along Γ–A line, AH/AH' lines and at the H/H' points, respectively. (d) The Fermi surface of $IrSi_4$.

The synergy between intrinsic topological band features (e.g., symmetry-protected Weyl points) and the superconducting order parameter can induce mixed-parity pairing and disorder-robust topological surface states. Combined with the nonreciprocal responses enabled by broken inversion symmetry, such as the superconducting diode effect, this interplay opens pathways toward low-dissipation chiral superconducting devices. Moreover, topological Fermi surfaces and chiral spin textures offer a platform for exotic spin-triplet pairing, deepening our understanding of unconventional superconductivity driven by strong spin-orbit coupling in correlated systems.

## CONCLUSIONS

We report the high-pressure synthesis and characterization of chiral-structured superconductors $RhGe_4$, $IrGe_4$, and $IrSi_4$. Their chiral crystal structures (space group $P3_121$) were unambiguously determined by combining single-crystal and powder XRD together with atomic-resolution STEM. All three compounds exhibit type-II superconductivity with $T_c$ up to ~ 2.6 K and $IrGe_4$ shows clear evidence of mixed $s$+$p$-wave pairing, enhanced by strong spin–orbit coupling. First-principles calculations further reveal symmetry-protected Weyl points, indicating the coexistence of topological band features and superconductivity in a single chiral lattice. This work establishes chiral-structured superconductors as a promising platform to explore the interplay among noncentrosymmetric crystal symmetry, spin–orbit coupling, mixed-parity pairing, and topological states—opening new avenues toward unconventional superconductivity and topological quantum phenomena.

## ASSOCIATED CONTENT

### Supporting Information

The Supporting Information is available free of charge on the ACS Publications website at DOI. It includes the details of structure refinement and the analysis of the upper critical field, the specific heat, the TDO, and the superconducting gap symmetry.

## AUTHOR INFORMATION

### Corresponding Author

*mayh2@shanghaitech.edu.cn (YHM), *xxwu@itp.ac.cn (XXW), *xgwan@nju.edu.cn (XGW), *hqyuan@zju.edu.cn (HQY), *guoyf@shanghaitech.edu.cn (YFG).

### Author Contributions

†These authors contributed equally.

### Notes

The authors declare no competing financial interest.

## ACKNOWLEDGEMENT

The authors acknowledge the National Key R&D Program of China (Grants No. 2023YFA1406100 and 2024YFA1408400). Y.F.G. acknowledges the support by the Science and Technology Commission of Shanghai Municipality (25DZ3008200 and YDZX20253100001002) and the open research fund of Beijing National Laboratory for Condensed Matter Physics (2023BNLCMPKF002). X.G.W. is supported by the New Cornerstone Science Foundation. X.W. is supported by the National Key R&D Program of China (Grant No. 2023YFA1407300) and the National Natural Science Foundation of China (Grants No. 12574151, No. 12447103, and No. 12447101). H.Q.Y. acknowledges the supports by the National Key R&D Program of China (No. 2022YFA1402200), the National Science Foundation of China (No. W2511006, No.12494592), and the New Cornerstone Science Foundation. Part of this work was carried out at the Synergetic Extreme Condition User Facility (SECUF, https://cstr.cn/31123.02.SECUF). The authors also thank the support from Analytical Instrumentation Center (#SPST-AIC10112914) and the Double First-Class Initiative Fund of ShanghaiTech University.

**TOC graphic**

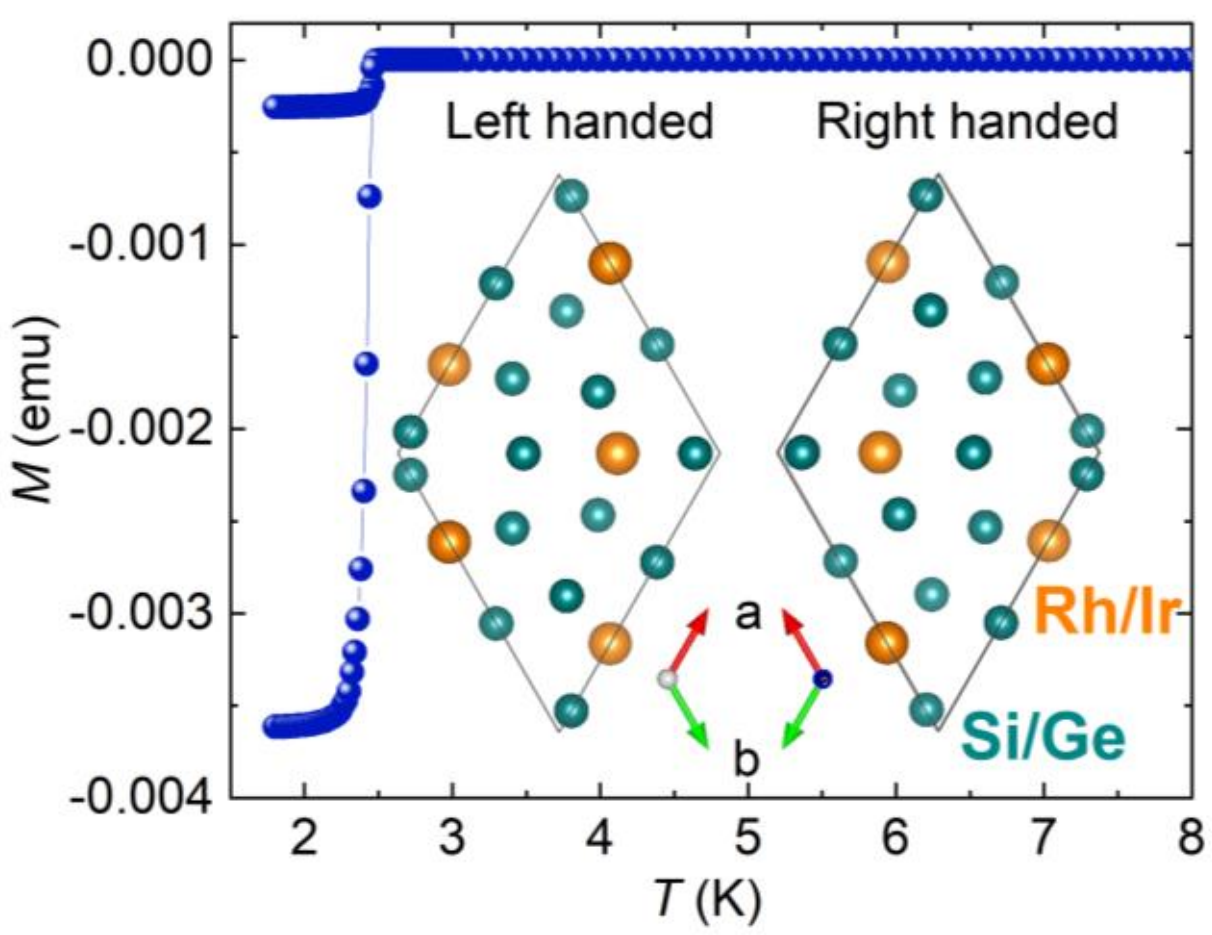

0.000
-0.001
-0.002
-0.003
-0.004
M (emu)
2
3
4
5
6
7
8
T (K)
Left handed
Right handed
a
b
Rh/Ir
Si/Ge

# SI

## 1. Chemical composition characterization of $RhGe_4$ and $IrGe_4$

Single crystals of $RhGe_4$ and $IrGe_4$ are shown in the insets of Figs. S1(a) and S1(b), respectively. Their stoichiometric ratios were determined by energy-dispersive X-ray spectroscopy (EDX). The average compositions measured at multiple points on each sample (Figs. S1(c) and S1(d)) confirm good stoichiometry, with the corresponding atomic ratios presented in Figs. S1(a) and S1(b).

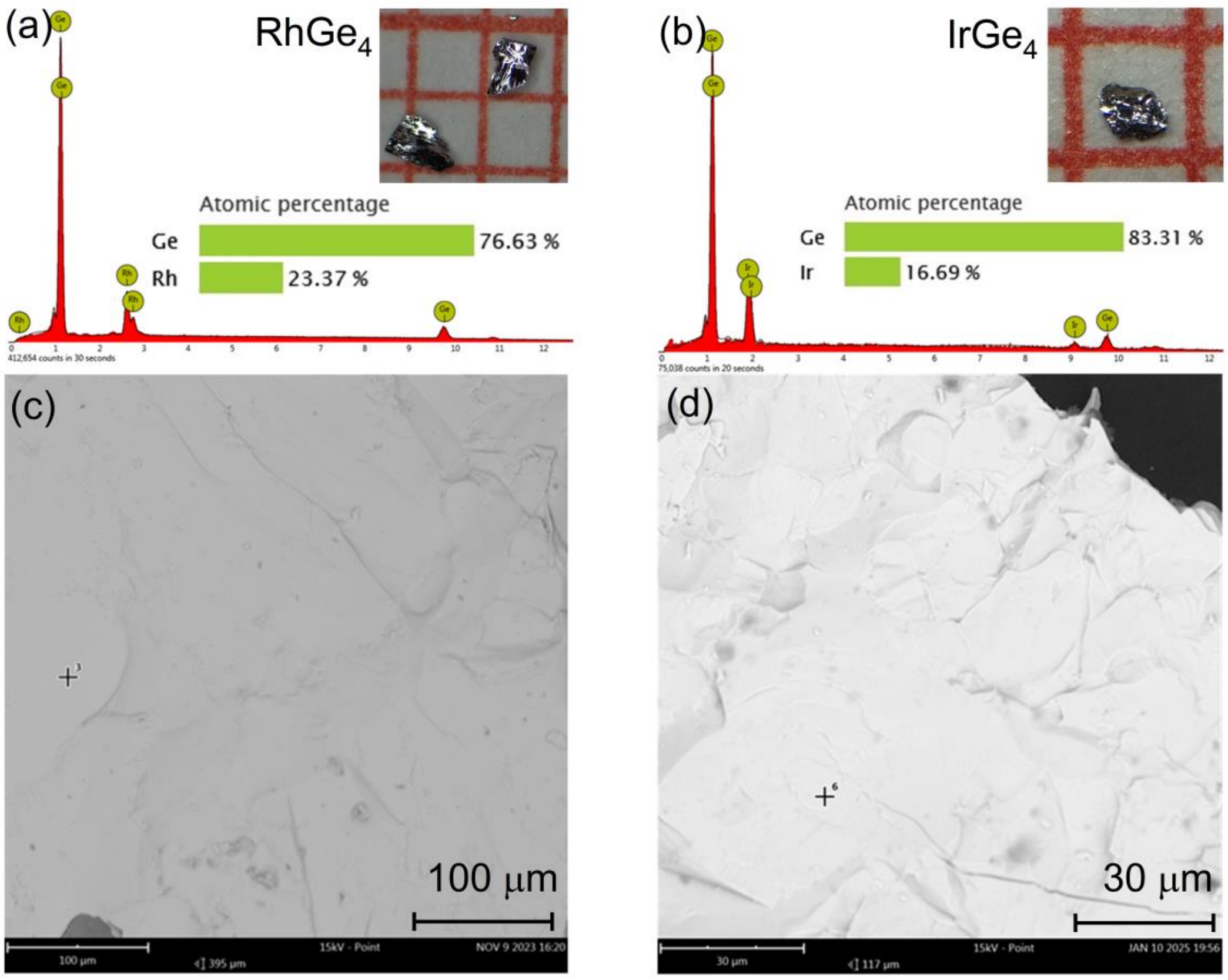

**Fig. S1**. (a) and (b) Optical pictures and EDX spectra of $RhGe_4$ and $IrGe_4$ single crystals prepared in the present work. (c) and (d) The SEM images of the present $RhGe_4$ and $IrGe_4$ single crystals.

## 2. Structure refinement results of $RhGe_4$, $IrGe_4$ and $IrSi_4$

**Table S1.** Crystal data and structure refinement for $RhGe_4$ at 290 K

| Empirical formula | $RhGe_4$ |
|---|---|
| Formula weight | 393.47 |
| Temperature | 290.0 K |
| Crystal system | Trigonal |
| Space group | $P3_121$ |
| Unit-cell dimens | $a$ = 6.1906(5) Å, $\alpha$ = 90° |
| | $b$ = 6.1906(5) Å, $\beta$ = 90° |
| | $c$ = 7.8082(7) Å, $\gamma$ = 120° |
| Volume $Z$ | 259.15(5) Å$^3$, 2 |
| Density (calcd) | 7.560 g/cm$^3$ |
| Absorp coeff | 38.793 mm$^{-1}$ |
| $F$ (000) | 519.0 |
| Crystal size (mm$^3$) | 0.01 × 0.05 × 0.01 |
| Radiation | Mo $K_\alpha$ ($\lambda$ = 0.71073 Å) |
| 2$\theta$ range for data collection | 7.602° to 68.874° |
| Index ranges | $-9 \leq h \leq 8$, $-8 \leq k \leq 9$, $-11 \leq l \leq 11$ |
| Reflections collected | 9180 |
| Independent reflections | 622 [$R_{int}$ = 0.1053, $R_{sigma}$ = 0.0555] |
| Data/restraints/parameters | 622/0/25 |
| Goodness-of-fit on $F^2$ | 1.107 |
| Final $R$ indexes [I >= 2σ (I)] | $R_1$ = 0.0341, $wR_2$ = 0.0610 |
| Final $R$ indexes (all data) | $R_1$ = 0.0476, $wR_2$ = 0.0639 |
| Largest diff.peak/hole / $e$ (Å$^{-3}$) | 1.87/-2.18 |
| Flack parameter | 0.044(16) |

**Table S2.** Crystal data and structure refinement for $IrGe_4$ at 290 K

| Empirical formula | $IrGe_4$ |
|---|---|
| Formula weight | 482.56 |
| Temperature | 290.0 K |
| Crystal system | Trigonal |
| Space group | $P3_121$ |
| Unit-cell dimens | $a$ = 6.1974(13) Å, $\alpha$ = 90° |
| | $b$ = 6.1974(13) Å, $\beta$ = 90° |
| | $c$ = 7.7649(18) Å, $\gamma$ = 120° |
| Volume $Z$ | 258.28(12) Å$^3$, 3 |
| Density (calcd) | 9.308 g/cm$^3$ |
| Absorp coeff | 72.736 mm$^{-1}$ |
| $F$ (000) | 615.0 |
| Crystal size (mm$^3$) | 0.01 × 0.05 × 0.01 |
| Radiation | Mo $K_\alpha$ ($\lambda$ = 0.71073 Å) |
| $2\theta$ range for data collection | 7.592° to 56.686° |
| Index ranges | $-8 \leq h \leq 8$, $-8 \leq k \leq 8$, $-10 \leq l \leq 10$ |
| Reflections collected | 2726 |
| Independent reflections | 437 [$R_{int}$ = 0.1233, $R_{sigma}$ = 0.0706] |
| Data/restraints/parameters | 437/0/26 |
| Goodness-of-fit on $F^2$ | 1.076 |
| Final $R$ indexes [I >= 2σ (I)] | $R_1$ = 0.0402, $wR_2$ = 0.0961 |
| Final $R$ indexes (all data) | $R_1$ = 0.0446, $wR_2$ = 0.0982 |
| Largest diff.peak/hole / $e$ (Å$^{-3}$) | 3.09/-2.41 |
| Flack parameter | 0.06(5) |

**Table S3.** Structure refinement for sample $IrSi_4$ prepared in the present work at 290 K

| $IrSi_4$, SG: $P3_121$, $a$ = 5.9629(3) Å and $c$ = 7.4777(3) Å, $R_p$ = 9.14%, $R_{wp}$ = 12.02%, weight percent: ~ 90% | | | | |
|---|---|---|---|---|
| Atom | WP | $x$ | $y$ | $z$ |
| Ir | 3a | 0.3147 | 0 | 0.3333 |
| Si(1) | 3b | 0.0701 | 0 | 0.8333 |
| Si(2) | 3b | 0.6073 | 0 | 0.8333 |
| Si(3) | 6c | 0.4866 | -0.2704 | 0.2566 |
| $IrSi_3$, SG: $P6_3mc$, $a$ = 4.3956(1) Å and $c$ = 6.6913(7) Å, $R_p$ = 9.14%, $R_{wp}$ = 12.02%, weight percent: ~ 10% | | | | |
| Atom | WP | $x$ | $y$ | $z$ |
| Ir | 2b | 1/3 | 2/3 | 0 |
| Si | 6c | 0.818 | 0.182 | 0.167 |

**Table S4.** Selected bond distances (Å) in $RhGe_4$, $IrGe_4$ and $IrSi_4$

| Bond | distance | Bond | distance | Bond | distance |
|---|---|---|---|---|---|
| Rh-Ge1 | 2.55(7) | Ir-Ge1 | 2.5498(19) | Ir-Si1 | 2.46(8) |
| Rh-Ge1 | 2.55(7) | Ir-Ge1 | 2.5498(19) | Ir-Si1 | 2.46(8) |
| Rh-Ge2 | 2.59(1) | Ir-Ge2 | 2.595(2) | Ir-Si2 | 2.48(5) |
| Rh-Ge2 | 2.59(1) | Ir-Ge2 | 2.595(2) | Ir-Si2 | 2.48(5) |
| Rh-Ge3 | 2.50(7) | Ir-Ge3 | 2.595(2) | Ir-Si3 | 2.38(5) |
| Rh-Ge3 | 2.50(7) | Ir-Ge3 | 2.506(3) | Ir-Si3 | 2.38(5) |
| Rh-Ge3 | 2.51(1) | Ir-Ge3 | 2.506(3) | Ir-Si3 | 2.41(6) |
| Rh-Ge3 | 2.51(1) | Ir-Ge3 | 2.525(3) | Ir-Si3 | 2.41(6) |
| Ge1-Ge2 | 3.33(6) | Ge1-Ge2 | 2.853(5) | Si1-Si2 | 3.16(3) |
| Ge1-Ge3 | 2.57(1) | Ge1-Ge3 | 2.555(3) | Si1-Si3 | 2.43(5) |
| Ge2-Ge3 | 2.80(5) | Ge2-Ge3 | 2.824(3) | Si2-Si3 | 2.69(1) |

## 3. STEM characterizations of $IrGe_4$

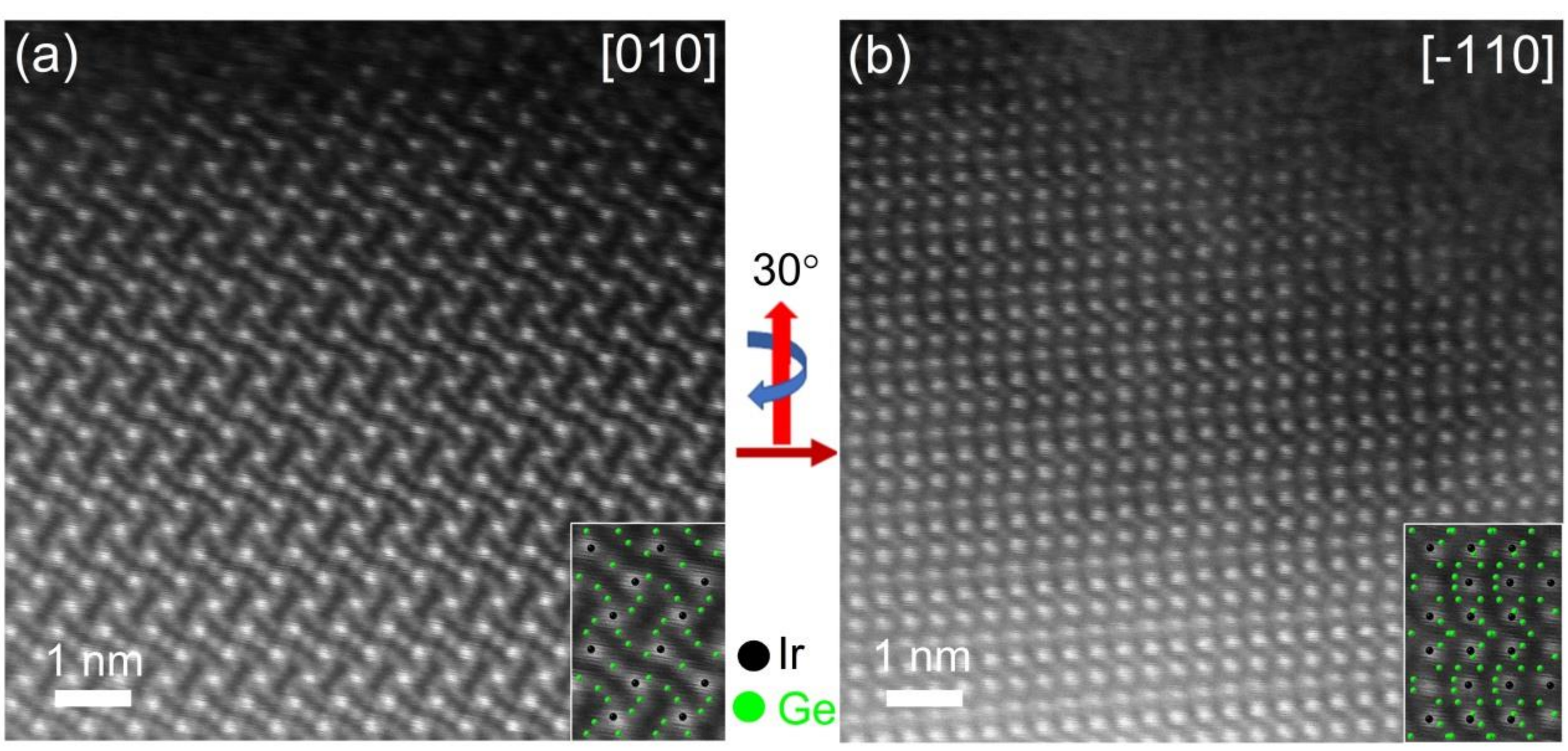

**Fig. S2.** The experimental images along two zone axes consistently match with the simulated projections for the $P3_121$ structure.

## 4. The $\rho(T)$, $M(T)$ and $C(T)$ properties of $RhGe_4$, $IrGe_4$ and $IrSi_4$

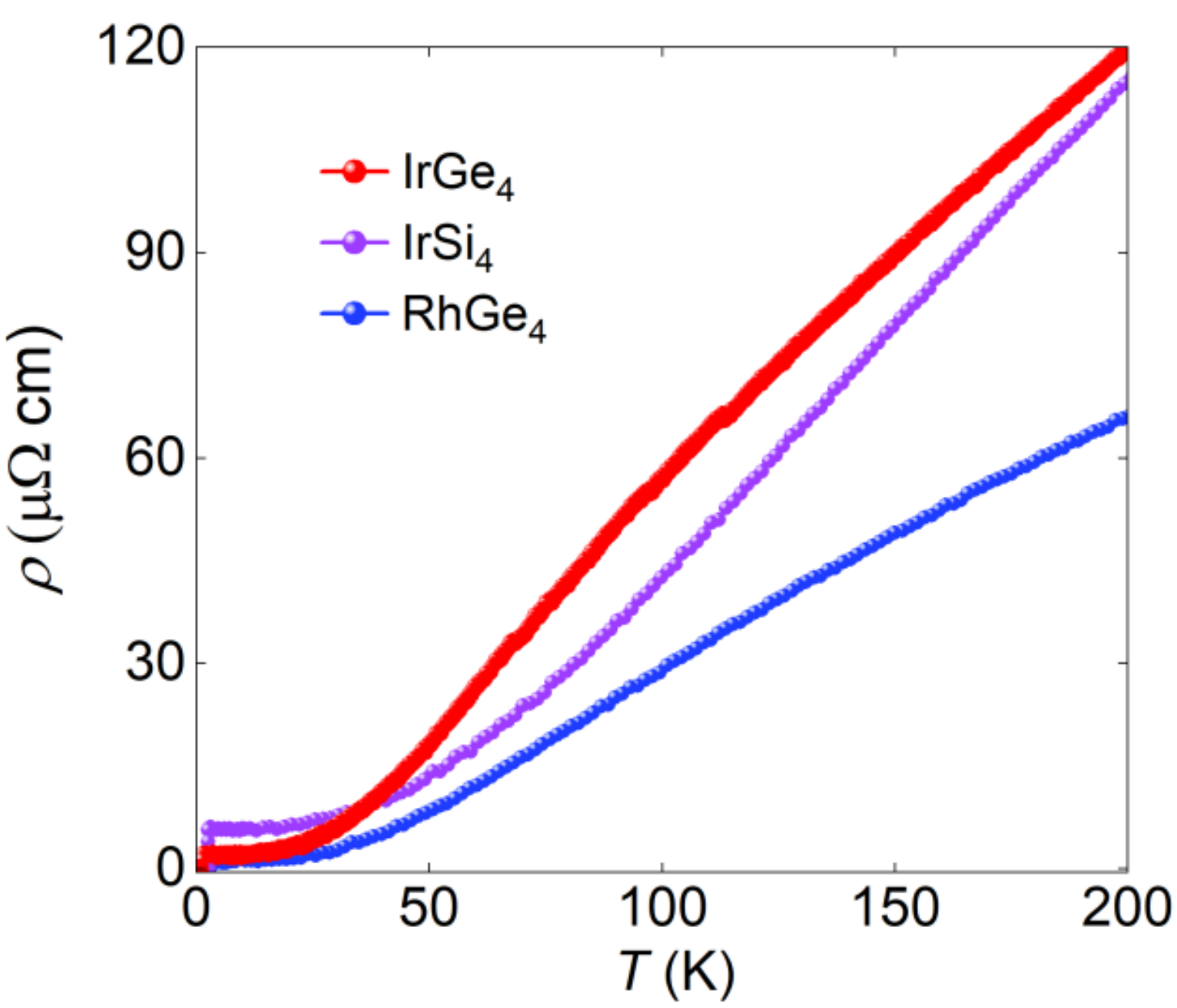

**Fig. S3.** Temperature-dependent resistivity $\rho(T)$ of $RhGe_4$, $IrGe_4$ and $IrSi_4$ during 1.8-200 K.

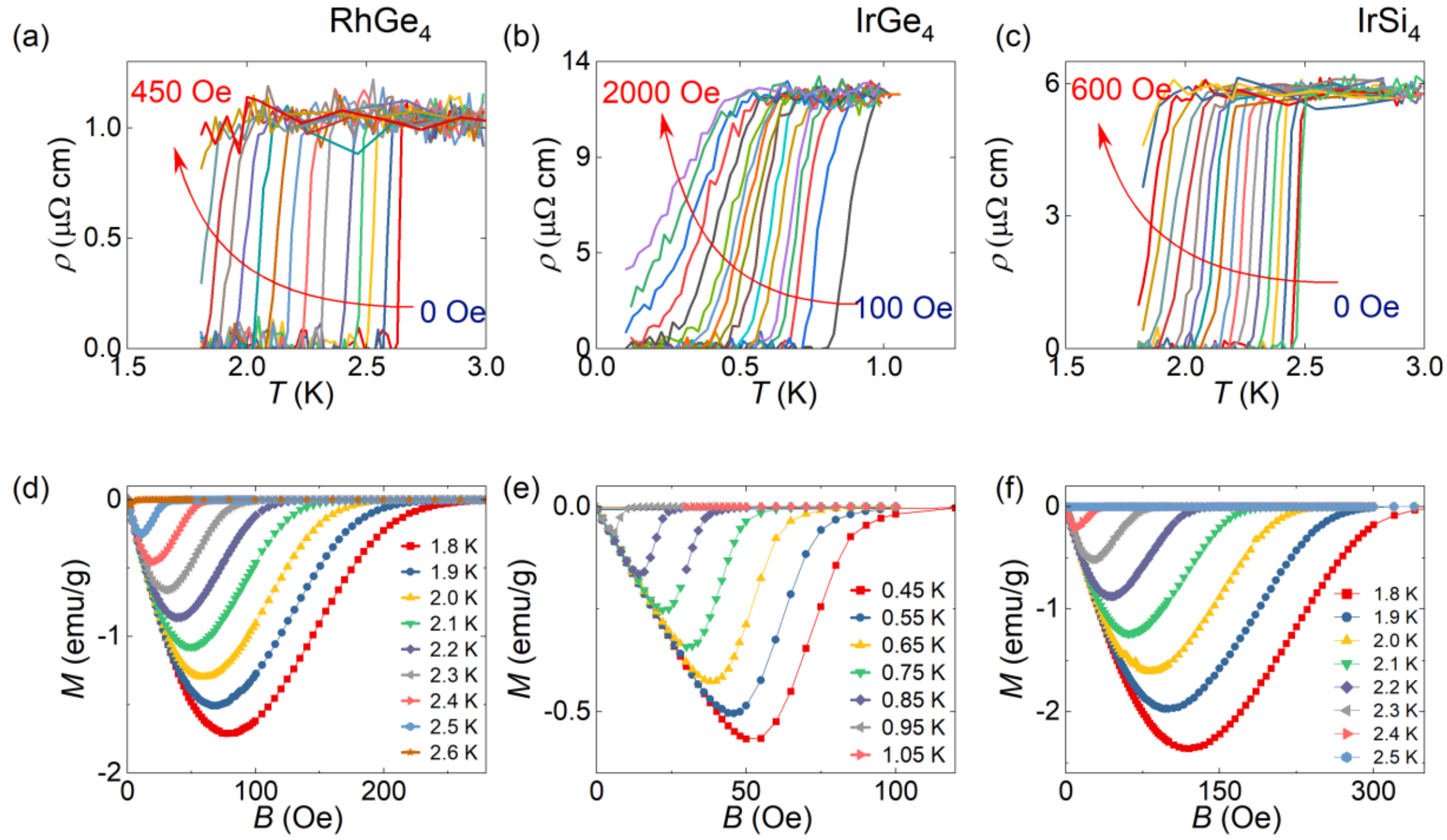

**Fig. S4.** (a)-(c) show the $\rho(T)$ data of $RhGe_4$, $IrGe_4$ and $IrSi_4$ under various applied magnetic fields. (d)-(f) show the magnetic field dependences of the magnetization for $RhGe_4$, $IrGe_4$ and $IrSi_4$ taken at several different temperatures.

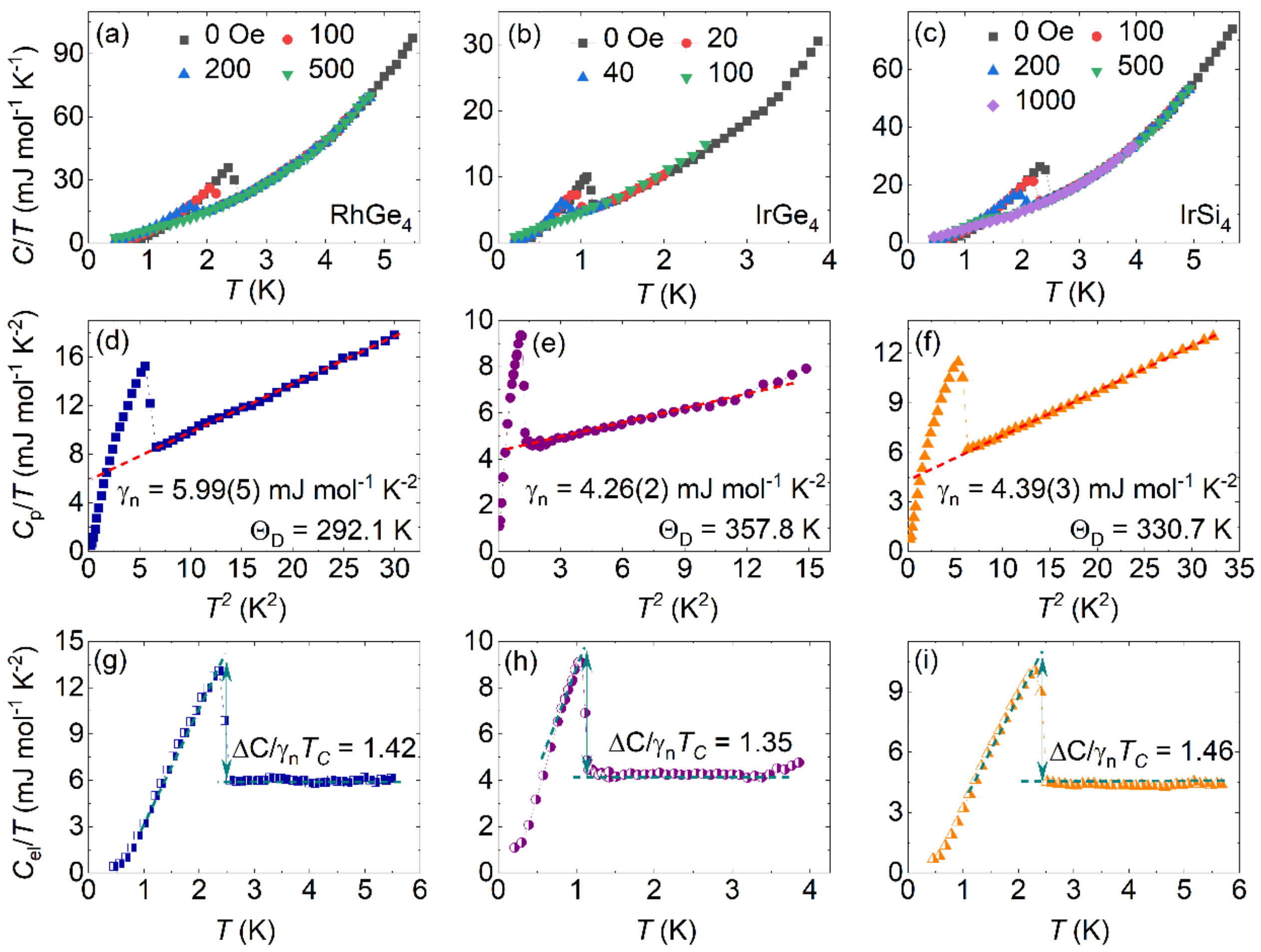

**Fig. S5.** The specific heat jump in zero magnetic field at low temperatures under applied magnetic fields ((a)-(c)), $C_p/T$ versus $T^2$ ((d)-(f)) as well as the estimation of normalized specific heat jump ($\Delta C/\gamma_n T_c$) ((g)-(i)) for $RhGe_4$, $IrGe_4$ and $IrSi_4$, respectively.

## 5. Specific heat data analysis of $RhGe_4$, $IrGe_4$, and $IrSi_4$

The $C_{el}/T$ data of $RhGe_4$, $IrGe_4$, and $IrSi_4$ were analyzed using several gap-function models, as shown in Figs. 4(a)–4(c). Following the analysis method reported for the noncentrosymmetric superconductor CaPtAs [S1], the superconducting electronic entropy was then calculated using the BCS formalism:

$$S(T) = -\left(\frac{6\gamma}{\pi^2 k_B}\right)\int_0^\infty [f\ lnf\ + (1-f)ln(1-f)]\, d\varepsilon,$$

where $f$ = $[1+ \exp(E/k_BT)]^{-1}$ is the Fermi distribution function and $E = \left(\varepsilon^2 + \Delta_k^2(T)\right)^{1/2}$ is the quasiparticle excitation energy. $E$ is the electron energy measured from the Fermi level. Here, $\Delta_k(T)$ is the temperature-and angle dependent gap function, The temperature dependence of the gap was assumed to follow $\Delta(T) = \Delta_0(0)\, tanh\{1.82[1.018(T_c/T - 1)]^{0.51}\}$, where $\Delta_0(0)$ is gap value in the zero-temperature limit. The electronic specific heat was obtained from $C_e(T) = T(dS/dT)$. Different superconducting gap symmetries were considered by introducing different angular gap functions, including isotropic *s*-wave ($g_k$ = 1), *p*-wave with point nodes ($g_k$ = sin$\theta$), and *d*-wave with line nodes ($g_k$ = cos2$\phi$). The gap function can be written as $\Delta_k(T) = \Delta_0(T)g_k$. In case of multiple gap models, the electronic specific heat can be modeled with:

$$C_e(T) = \alpha C_{e,1}^{\Delta^1}(T) + (1-\alpha)C_{e,2}^{\Delta^2}(T),$$

where $\Delta^1$ and $\Delta^2$ represent two superconducting gaps and $\alpha$ is the relative weight of the first gap component. By comparing the fitting quality of different models, the superconducting gap structure and possible nodal behavior can be evaluated.

**Table S5.** The detailed specific heat fitting results of $IrGe_4$, $RhGe_4$ and $IrSi_4$

**$IrGe_4$**

| | Model | $\alpha$, 1-$\alpha$ | $\Delta$ (meV) | Adj. $R^2$ |
|---|---|---|---|---|
| i | *s* wave | 1, 0 | 0.1584 | 0.9849 |
| ii | *p* wave | 1, 0 | 0.2343 | 0.9948 |
| iii | *d* wave | 1, 0 | 0.2991 | 0.9885 |
| iv | *s* + *s* | 0.8487, 0.1513 | 0.2116, 0.0397 | 0.9994 |
| v | *s* + *p* | 0.8430, 0.1570 | 0.2075, 0.2419 | 0.9995 |
| vi | *s* + *d* | 0.4388, 0.5612 | 0.2297, 0.2488 | 0.9985 |

$C_e = \alpha\, C_{e,1} + (1 - \alpha)\, C_{e,2}$, $\gamma_n = 4.14(6)$ mJ mol$^{-1}$ K$^{-2}$, $\beta = 0.23(1)$ mJ mol$^{-1}$ K$^{-4}$.

**$RhGe_4$**

| | Model | $\alpha$, 1-$\alpha$ | Δ (meV) | Adj. $R^2$ |
|---|---|---|---|---|
| i | *s* wave | 1, 0 | 0.3668 | 0.9989 |
| ii | *p* wave | 1, 0 | 0.4519 | 0.9924 |
| iii | *d* wave | 1, 0 | 0.5116 | 0.9642 |
| iv | *s* + *s* | 0.0357, 0.9643 | 0.0045, 0.3716 | 0.9990 |
| v | *s* + *p* | 0.8870, 0.1130 | 0.3819, 0.2588 | 0.9980 |
| vi | *s* + *d* | 0.9004, 0.0996 | 0.3806, 0.2641 | 0.9981 |

$C_e = \alpha\, C_{e,1} + (1 - \alpha)\, C_{e,2}$, $\gamma_n = 5.99(5)$ mJ mol$^{-1}$ K$^{-2}$, $\beta = 0.39(1)$ mJ mol$^{-1}$ K$^{-4}$.

**$IrSi_4$**

| | Model | $\alpha$,1-$\alpha$ | Δ (meV) | Adj. $R^2$ |
|---|---|---|---|---|
| i | *s* wave | 1, 0 | 0.3467 | 0.9951 |
| ii | *p* wave | 1, 0 | 0.4349 | 0.9911 |
| iii | *d* wave | 1, 0 | 0.5074 | 0.9646 |
| iv | *s* + *s* | 0.9044, 0.0956 | 0.3696, 0.0479 | 0.9988 |
| v | *s* + *p* | 0.9030, 0.0970 | 0.3698, 0.0546 | 0.9988 |
| vi | *s* + *d* | 0.8614, 0.1386 | 0.3576, 0.3971 | 0.9954 |

$C_e = \alpha\, C_{e,1} + (1 - \alpha)\, C_{e,2}$, $\gamma_n = 4.39(3)$ mJ mol$^{-1}$ K$^{-2}$, $\beta = 0.27(2)$ mJ mol$^{-1}$ K$^{-4}$.

## 6. Penetration depth and superfluid density analysis of $RhGe_4$ and $IrGe_4$

For $RhGe_4$, the data over the entire temperature range are well described by the single-gap *s*-wave model, with an optimal fitting parameter $\Delta(0) = 1.67\ k_BT_c$, whereas the *p*-wave model exhibits significant deviation at low temperatures. However, the gap value obtained from the single-gap *s*-wave fit is considerably larger than that from the low-temperature penetration depth fit ($1.11\ k_BT_c$), indicating that although the single-gap *s*-wave model can mathematically describe the overall curve, it may not be the optimal description in the low-temperature limit. To further examine the possibility of multigap superconductivity, the two-gap *s*-wave model is employed to refit the superfluid density. In this model [S2], the total superfluid density $\rho_s(T)$ can be expressed as $\rho_s(T) = x\rho_{s1} + (1-x)\rho_{s2}$, where $\rho_{s1}$ and $\rho_{s2}$ are the superfluid densities calculated from $\Delta_1$ and $\Delta_2$ according to $\rho_s(T) = 1 + 2 < \int_{\Delta_k}^{+\infty} \frac{EdE}{\sqrt{E^2-\Delta_k^2}} \frac{\partial f}{\partial E} >_{FS}$, respectively, and $x$ is the relative weight of $\rho_{s1}$ ($0 < x < 1$). The fitting results reveal that the small gap in $RhGe_4$ accounts for less than 5% weight. Such a small weight cannot serve as conclusive evidence for an independent small gap. Nevertheless, the gap magnitude from the low-temperature $\Delta\lambda(T)$ analysis is considerably smaller than the weak-coupling BCS value, and also much smaller than that from the single-gap $s$-wave fit to the superfluid density. Taken together with the limited small-gap weight, $RhGe_4$ is identified as a nodeless superconductor that may exhibit weak anisotropy or multiband features.

For $IrGe_4$, the single-gap $s$-wave model shows significant deviation from the experimental data even with the optimal fitting parameter $\Delta(0) = 2.3\ k_BT_c$. Fitting with the two-gap $s$-wave model gives $\Delta_1(0) = 1.27\ k_BT_c$, $\Delta_2(0) = 2.85\ k_BT_c$, with a small-gap weight of 0.26. The value of the small gap is slightly larger than that expected from the low-temperature superfluid density fit ($0.71\ k_BT_c$), indicating a minor discrepancy. In contrast, the $p$-wave model yields a fit that is much closer to the data. Although the low-temperature penetration depth change $\Delta\lambda(T)$ of $IrGe_4$ does not provide very clear evidence for point nodes ($p$-wave), the overall behavior of its superfluid density is in good agreement with the $p$-wave model. Therefore, the superfluid density of $IrGe_4$ is further analyzed using a model based on the two spin-split bands [S3]. For $IrGe_4$ (space group: $P3_121$, NO. 152, point group $D_3$), $\mathbf{g}_k \propto (a_{xx}k_x, a_{xx}k_y, a_{zz}k_z)$[S4]. Given the current lack of detailed band structure calculations for $IrGe_4$, an effective model under the spherical Fermi surface approximation, $|\mathbf{g}_k| = \sin\theta$, is adopted to qualitatively analyze its superfluid density behavior.

**Table S6.** The detailed superfluid density fitting results of $IrGe_4$ and $RhGe_4$

**$IrGe_4$**

| | Model | $\alpha$, 1-$\alpha$ | $\Delta$ (meV) | Adj. $R^2$ |
|---|---|---|---|---|
| i | $s$ wave | 1, 0 | 0.217 | 0.9897 |
| ii | $p$ wave | 1, 0 | 0.297 | 0.9991 |
| iii | $s + s$ | 0.26, 0.74 | 0.120,0.175 | 0.9994 |
| iv | $s + p$ | 0.29, 0.71 | 0.168, 0.341 | 0.9997 |

**$RhGe_4$**

| | Model | $\alpha$, 1-$\alpha$ | Δ (meV) | Adj. $R^2$ |
|---|---|---|---|---|
| i | *s* wave | 1, 0 | 0.373 | 0.9989 |
| ii | *p* wave | 1, 0 | 0.488 | 0.9953 |
| iii | *s* + *s* | 0.04, 0.96 | 0.246,0.385 | 0.9987 |
| iv | *s* + *p* | 1, 0 | 0.373 | 0.9989 |

**Table S7.** Summary of the literature reported and present chiral crystal structure superconductors

| formula | $T_c$ (K) | $H_{c2}$ (T) | Space group | Ref. |
|---|---|---|---|---|
| $\alpha$-BiPd | 3.8 | 0.70 | $P2_1$ (No. 4) | [S5] |
| UIr | $2.6^p$ | $0.0258^p$ | $P2_1$ (No. 4) | [S6] |
| $NbRh_2B_2$ | 7.6 | 18.0(4) | $P3_1$ (No. 144) | [S7] |
| $TaRh_2B_2$ | 5.8 | 11.7(4) | $P3_1$ (No. 144) | [S8] |
| $RhGe_4$ | 2.6 | 0.10 | $P3_121$ (No. 152) | [S8,S9,PW] |
| $IrGe_4$ | 1.1 | 0.30 | $P3_121$ (No. 152) | [S9,PW] |
| $IrSi_4$ | 2.5 | 0.18 | $P3_121$ (No. 152) | PW |
| $(Pt_{0.2}Ir_{0.8})_3Zr_5$ | 2.2 | 2.59 | $P6_122$ (No. 178) | [S10] |
| $NbSi_2$ | 0.13 | NP | $P6_222$ (No. 180) | [S11] |
| $TaSi_2$ | 0.353 | 0.0298 | $P6_222$ (No. 180) | [S12] |
| $NbGe_2$ | 2.1 | 0.1 | $P6_222$ (No. 180) | [S13] |
| $TaGe_2$ | 2.3 | NP | $P6_222$ (No. 180) | [S14] |
| RhGe | 4.3 | NP | $P2_13$ (No. 198) | [S8] |
| BeAu | 3.3 | 0.03 | $P2_13$ (No. 198) | [S15] |
| $Li_2Pd_3B$. | 7.0 | 4.8 | $P4_132$ (No. 213)) | [S16] |
| $Li_2Pt_3B$ | 2.4 | 1.9 | $P4_132$ (No. 213)) | [S17] |
| $Mo_3Al_2C$ | 9.2 | 18.2 | $P4_132$ (No. 213) | [S18] |
| $Mo_7Re_{13}B$ | 8.3 | 15.4 | $P4_132$ (No. 213) | [S19] |
| $Mo_7Re_{13}C$ | 8.1 | 14.8 | $P4_132$ (No. 213) | [S19] |
| $Mo_3Rh_2N$ | 4.6 | 7.32 | $P4_132$ (No. 213) | [S20] |
| $W_7Re_{13}B$ | 7.1 | 11.4 | $P4_132$ (No. 213) | [S19] |
| $W_7Re_{13}C$ | 7.3 | 12.6 | $P4_132$ (No. 213) | [S19] |
| $W_4IrC$ | 4.8 | 8.3 | $P4_132$ (No. 213) | [S21] |
| $Cr_2Re_3B$ | 4.8 | 10 | $P4_123$ (No. 213) | [S22] |

*P* = 2.6 GPa, Present work (PW), Not been reported (NP)